\documentclass{article}

\usepackage{PRIMEarxiv}

\usepackage[utf8]{inputenc} % allow utf-8 input
\usepackage[T1]{fontenc}    % use 8-bit T1 fonts
\usepackage{hyperref}       % hyperlinks
\usepackage{url}            % simple URL typesetting
\usepackage{booktabs}       % professional-quality tables
\usepackage{amsfonts}       % blackboard math symbols
\usepackage{nicefrac}       % compact symbols for 1/2, etc.
\usepackage{microtype}      % microtypography
\usepackage{lipsum}
\usepackage{fancyhdr}       % header
\usepackage{graphicx}       % graphics
\graphicspath{{media/}}     % organize your images and other figures under media/ folder
\usepackage{amssymb}
\usepackage{subfig}
\newtheorem{thm}{\textbf{Theorem}}

\title{Market Making and Pricing of Financial Derivatives based on Road Travel Time
\thanks{\textit{\underline{Citation}}: 
\textbf{Ke Wan, Alain Kornhauser: Market Making and Pricing of Financial Derivatives based on Road Travel Time}} 
}

\author{
  Ke Wan \\
  Dr. \\
  Princeton University \\
  Princeton\\
  \texttt{kwan@alumni.princeton.edu} \\
   \And
  Alain Kornhauser \\
  Professor \\
  Princeton University \\
  Princeton\\
  \texttt{alaink@princeton.edu} \\
}

\begin{document}
\maketitle

\begin{abstract}
  
  Travel time derivatives are financial instruments that derive their value from road travel times, serving as an underlying asset that cannot be directly traded. Within the transportation domain, these derivatives are proposed as a more comprehensive approach to value pricing. They enable road pricing based not only on the level of travel time but also its volatility. In the financial market, travel time derivatives are introduced as innovative hedging instruments to mitigate market risk, particularly in light of recent stress experienced by the crypto market and traditional banking sector.

  The paper focuses on three main aspects: (1) the motivation behind the introduction of these derivatives, driven by the demand for hedging; (2) exploring the potential market for these instruments; and (3) delving into the product design and pricing schemes associated with them. The pricing schemes are devised by utilizing real-time travel time data captured by sensors. These data are modeled using Ornstein-Uhlenbeck processes and, more broadly, continuous time autoregressive moving average (CARMA) models. The calibration of these models is achieved through a hidden factor model, which describes the dynamics of travel time processes. The risk-neutral pricing principle is then employed to determine the prices of the derivatives, employing well-designed procedures to identify the market value of risk.

\end{abstract}

\subsection{Keywords}
Travel time derivatives, non-tradable asset, financial derivatives,
asset pricing, travel time forecast, time series, CARMA process,
hidden factor models.

\section{Initiation and necessity analysis}

%This paper introduces the concept of a travel time derivative
%%as an
%%alternative approach to congestion pricing, sometimes referred to as
%%value pricing, for use by transportation facilities.

This paper introduces the concept of a travel time derivative as an
alternative approach to congestion pricing, sometimes referred to as
value pricing, for use by transportation facilities.
 A travel time derivative is a useful financial product to (a) hedge against
transportation-related risk by users of the transportation facility,
(b) manage the demand for those facilities through derivatives-based
dynamic tolling, (c) contribute to risk mitigation through portfolio
diversification and (d) provide an additional source of revenue for
owners of transportation facilities. In this paper, potential market
participants are analyzed first, and then major products designed
for a travel time derivatives market are presented. Alternative
models for describing underlying travel time changes are discussed,
together with corresponding pricing methods.

\subsection{Derivatives and weather derivatives as hedging tools}

Derivatives are financial instruments whose prices are derived from
the value of something else, known as the underlying asset. The
major types of derivatives are forwards, futures, options, and swaps
\cite{john2000options}. Any stochastic changing element that
generates changes in cash flow can serve as the underlying asset.
Therefore, the underlying element upon which a derivative is based
can be the price of an asset (e.g., commodities, equities [stock],
residential mortgages, commercial real estate, loans, bonds), the
value of an index (e.g., interest rates, exchange rates, stock
market indices, consumer price index [CPI]), or other items (e.g.,
temperature, precipitation). The underlying elements of derivatives
can be further classified as tradable and non-tradable. Most items
listed in the preceding paragraph as bases for derivatives can be
traded in a market and so are called tradable underlying assets. A
few others, such as temperature, precipitation, and travel time, are
non-tradable. This difference in the tradability of an underlying
asset triggers differences in market making mechanism, trading
strategy and pricing methods.

Derivatives based on non-tradable assets were first introduced in
1999, when the Chicago Mercantile Exchange introduced weather
futures contracts, the payoffs for which are based on average
temperatures at specified locations. According to
\cite{stewart2002derivative}, weather derivatives offer an
innovative hedging instrument to firms facing the possibility of
significant earnings declines or advances because of unpredictable
weather patterns. \cite{banks2002weather} analyzed participants and
roles in that futures market and found that weather derivatives act
as alternative and more flexible ways of insuring against weather
related risk. Industries subject to weather risk participate in the
buy/sell side of the market, while speculators, who trade purely for
profit, provide an important source of liquidity.

Weather derivatives provide insurance to farmers and agriculture
companies against bad weather and low crop output. The payoff for
one farmer who grows corn and buys a weather derivative contract is
as follows: When the weather is good, the insured benefits from
abundant corn output; when the weather is bad, the insured receives
extra compensation from the derivative to cover losses in corn
sales. In this way, the insured hedges risk. This risk-protection
mechanism is shown in Table \ref{tab:7-paticipants}. In the table,
$G$ represents the gain on the derivative, and $P$ represents the
premium that the farmer pays for the contract.

\begin{table}[ht]
\caption{Payoff of typical weather derivatives}
\label{tab:7-paticipants} \centering
\begin{tabular}{|c|c|c|}
\hline
Weather Condition&Corn production payoff&Derivative payoff\\
 \hline
Good &$G_{good}$&$-P$\\
Bad &$G_{bad}$&$G-P$\\
\hline
%\newnot{symbol:G}
%\newnot{symbol:Pprice}
\end{tabular}
\end{table}

As weather derivatives are used to hedge risk related to
temperature, precipitation, and other factors, the pricing of
weather derivatives should primarily be based upon prediction of
weather conditions. Based on accurate prediction of future weather
changes, the contract is priced using different methods other than
the Black-Scholes pricing model, considering the fact that weather
conditions are not traded in the market. These pricing methods are
not founded on dynamic hedging of the un-tradable underlying
instruments but on other more general pricing schemes such as risk
neutral pricing under incomplete market conditions, indifferent
pricing principles and etc, which are introduced in greater detail
in later sections.

\subsection{Road pricing for changing traffic patterns and generating revenue}

The concept of using road tolls as a means of reducing traffic congestion was first introduced by economist Arthur Pigou \cite{hau1992economic}. Tolls can serve as a revenue generator for road infrastructure financing, or as a transportation management tool to curb peak hour travel and related traffic congestion. A considerable amount of literature exists on the methodology of road tolls, and several studies are reviewed in this section.

Vovsha (2006) \cite{vovsha2006making} explored a wide range of possible modeling techniques for road pricing, from simplified sketch-planning tools for short-term revenue forecasting to comprehensive travel demand models for large-scale problems. The study emphasized models associated with advanced network simulation tools, such as dynamic traffic assignment and microsimulation, and advanced activity-based and tour-based demand models.

To evaluate the effectiveness of road pricing schemes, Lu (2008) \cite{lu2008bi} developed a bi-criterion dynamic user equilibrium (BDUE) model, which aims to capture users' path choices in response to time-varying toll charges. Zheng (2016) \cite{zheng2016time} proposed a time-dependent pricing scheme in which tolls are iteratively adjusted through a Proportional–Integral type feedback controller, based on the level of vehicular traffic congestion and traveler's behavioral adaptation to pricing costs.

Croci (2016) \cite{croci2016urban} suggested that road charges should be designed in a clear way, with charges high enough to induce travel behavior changes, coupled with an increase in public transport supply, and regularly updated. The results and benefits of these charges should be monitored and communicated to citizens in a timely manner. Lombardi (2021) \cite{lombardi2021model} reviewed recent research on the design, simulation, implementation, and evaluation of dynamic tolling schemes, covering control-based price definition rules and optimization-based algorithms. The study identified the main objectives of dynamic toll pricing, including maintaining free-flow conditions, minimizing travel times, and reducing externalities.

Hall (2021) \cite{hall2021can} found that the value of time, schedule inflexibility, and desired arrival time are the three important dimensions for evaluating the effects of adding optimal time-varying tolls. Additionally, the study found that adding tolls on half of the lanes of a highway yields a Pareto improvement.

Despite the various approaches to road pricing, existing methodologies have not closely linked the performance of the transportation system with the financial system's performance. Road pricing has primarily served as a policy tool rather than an interface that allows the financial market to impact and interact with the transportation system. This gap has led to the innovation of financial derivatives based on travel time.
\subsection{Travel time derivatives provide economic hedge against traffic delay}

Formally, a travel time derivative contract is a financial
instrument whose prices are derived from the value of travel time
measurements. The introduction of such a contract may bring several
benefits to the transportation and financial system, which are
addressed in this and following sections. Temperature changes at a
given location, on the one hand, and travel times along a given
path, on the other, both share similar stochastic patterns. Similar
to farmers, travelers could usefully be insured against the economic
costs of low-quality traffic service. This insurance can be
generated by using financial derivatives based on travel time.

%Different from traditional congestion pricing schemes,
Travel time derivatives not only impose a road toll but also provide
a corresponding payoff to travelers. For typical travelers, the
payoff is larger when the experienced travel time is high, which is
similar to insurance against bad quality of transportation service.

Here is an illustration of the payoff of a typical travel time
derivative contract. When a traveler experiences good traffic,
nothing needs to be paid except a premium $-P(T)$. The payoff is
defined as follows:

\begin{enumerate}
\item The payoff in the transportation system is good quality of service (QOS): $T_{good}$
\item The derivative payoff is $-P(T)$
%,as in Figure \ref{fig:7-payoffG}.
 \end{enumerate}

%\begin{figure}[htm]
%\begin{center}
%\includegraphics[angle=0, width=0.7\textwidth]{figures/Visio-TD-scenarionewgood}
%\caption{Travelers and travel time (quality of service) protection -
%good scenario: Good traffic conditions lead to low High Congestion
%Days. Therefore, HCD call option is out of money and there is no
%payoff to the traveler.}\label{fig:7-payoffG}
%\end{center}
%\end{figure}

When traveler experience bad traffic, a gain/compensation $G$ is
received while paying the premium $-P(T)$
%, as in Figure\ref{fig:7-payoffB}.
\begin{enumerate}
\item His payoff in the transportation system is bad QOS $T_{bad}$
\item His derivative payoff is $G-P(T)$, where $G$ is in proportion to the
experienced extra travel time from a predefined level $K$
 \end{enumerate}
%
%\begin{figure}[htm]
%\begin{center}
%\includegraphics[angle=0, width=0.7\textwidth]{figures/Visio-TD-scenarionewbad}
%\caption{Travelers and travel time(QOS) protection - bad scenario:
%Bad traffic conditions lead to high High Congestion Days. Therefore,
%HCD call option is in the money and there are payoff to the traveler
%to compensate his economic loss due to traffic
%delay.}\label{fig:7-payoffB}
%\end{center}
%\end{figure}

Based on the two scenario analyses above, comparisons of traditional
congestion pricing methods and travel time derivatives are given.
Traditionally, there are two categories of congestion pricing
schemes: static road toll and dynamic road toll (toll by time of day
and hence by congestion levels).

With static road tolls, the traveler pays a fixed premium/toll $P$
to use the road, as in Table \ref{tab:7-payoffT}. The toll is
constant no matter when the traveler enters the link. With dynamic
road tolls, the traveler pays a fixed amount P when the road has
less favorable conditions for travel (usually the prices are set
higher during rush hours) and pays nothing when the road has
favorable conditions for travel, as in Table \ref{tab:7-payoffD}.

With travel time derivatives the traveler's payment $P$ increases
continuously in tandem with expected traffic conditions, which
allows the traveler to benefit from compensation payoff $G$, which
is set in proportion to the quality of service received, as in Table
\ref{tab:7-payoffTD}. This comparison shows that the road tolls
charged through travel time derivatives are directly linked to
expected quality of service; the payoff can be set high enough such
that travelers will consider his financial income when making
routing choices. As a result, the travel pattern will change
accordingly.

%#hence, there are potentially more flexible and effective ways of

%#providing insurance against any economic cost exacted by poor
%#quality of traffic service in the future.

\begin{table}[ht]
\caption{Payoff of traditional road toll, $P$ denotes the toll
amount} \label{tab:7-payoffT} \centering
\begin{tabular}{|c|c|c|}
\hline
Traffic Condition&Traffic payoff&Derivative payoff\\
 \hline
Good traffic&$T_{good}$&$-P$\\
Bad traffic&$T_{bad}$&$-P$\\
\hline
\end{tabular}
\end{table}

\begin{table}[ht]
\caption{Payoff of dynamic congestion pricing, $P$ denotes the toll
amount} \label{tab:7-payoffD} \centering
\begin{tabular}{|c|c|c|}
\hline
Traffic Condition&Traffic payoff&Derivative payoff\\
 \hline
Rush Hour&$T_{good}$&$-P$\\
Other time&$T_{bad}$&$0$\\
\hline
\end{tabular}
\end{table}

\begin{table}[ht]
\caption{Payoff of a travel time derivatives. $P$ denotes its price,
$ET$ is the expected travel time and $K$ is strike price of the
derivative contract. The formula describes payoff as a function of
realized travel time and the strike price; the price of the
derivative contract is not zero but is calculated in proportion to
market participants' expectations regarding future traffic
conditions.} \label{tab:7-payoffTD} \centering
\begin{tabular}{|c|c|c|}
\hline
Traffic Condition&Traffic payoff&Derivative payoff\\
 \hline
Good traffic&$T_{good}$&$-P(T)$\\
Bad traffic&$T_{bad}$&$\alpha(T_{bad}-K)-P(T)$\\
\hline
%\newnot{symbol:Kstrike}
\end{tabular}
\end{table}

As flexible payoff functions can be defined, travel time derivatives
can provide a payoff according to travelers' experienced travel time
in future to reduce potential economic costs due to traffic delays.

Furthermore, travel time derivatives are useful for businesses whose
profits are related to traffic conditions. Firms can be
distinguished into two categories based on whether or not they
benefit from good traffic conditions.

For cargo transportation companies, profit are derived from daily
transportation service. If overall traffic conditions are bad, there
are more delays for trucks, therefore overall service to clients is
worse and operation costs are increased. The profits of the company
are reduced and it can lose competitive advantage in the
marketplace. With travel time derivatives, the company can invest in
travel time derivatives and hedge potential losses due to bad
traffic conditions in the future. If the overall traffic conditions
are good, the service is good and the firm can obtain potentially
increasing profits. The cost is just a premium which is used to
purchase derivative contracts, if overall traffic conditions are
poor the following year.

On the other hand, some firms would profit if traffic conditions are
worse, including toll road owners, public transportation companies,
companies offering alternative transportation services, etc. When
traffic conditions are good, fewer travelers will select toll roads,
therefore toll road owners tend to have less profit when overall
travel time is low, i.e., they are hurt by good traffic conditions.
Similarly, fewer people would select public transportation or
alternative transportation including trains over driving themselves
when traffic conditions are good, therefore related firms are also
hurt from good traffic conditions. Travel time derivatives provide
methods for them to hedge their risks when traffic conditions are
good.

As different businesses have different payoffs based on the
performance of traffic systems, they have incentives to hedge their
risk over the market. Different risk appetites between different
market participants lead to diversified trading activities. The
firms who do not profit from good traffic conditions can trade
against those that benefit from good traffic conditions. These
hedging activities provide strong incentives for introducing travel
time derivatives.

\subsection{Travel time derivatives are a flexible value pricing scheme which changes traveler's behavior}
%#% insurance # change behavior

Furthermore, because the price of travel time derivatives changes as
the predicted traffic conditions change, travel time derivatives can
be used to predict future travel time and change travelers' route
choice, by which their true time cost caused by traffic delays can
be reduced.%
 To enable  flexible protections, there are two kinds of payoff
functions for travel time derivatives, which differ in the time span
covered by underlying travel time measures.

\begin{enumerate}
\item The first type of travel time derivatives can be based on the
instantaneous travel time measures at given locations in the future.
As the graph shows, for some travel time derivatives of this type,
the market participants believe the economic loss due to high travel
time is going to be lower or less volatile if and only if the price
of the derivative contract is higher. Therefore, the price of travel
time derivatives indicates economic loss due to travel time in the
short term and travelers can select the paths with higher prices
when making routing decisions. This category of travel time
derivative is demonstrated in Figure \ref{fig:7PredictionEffect}.

\begin{figure}[ht]
\begin{center}
\includegraphics[angle=0, width=0.7\textwidth]{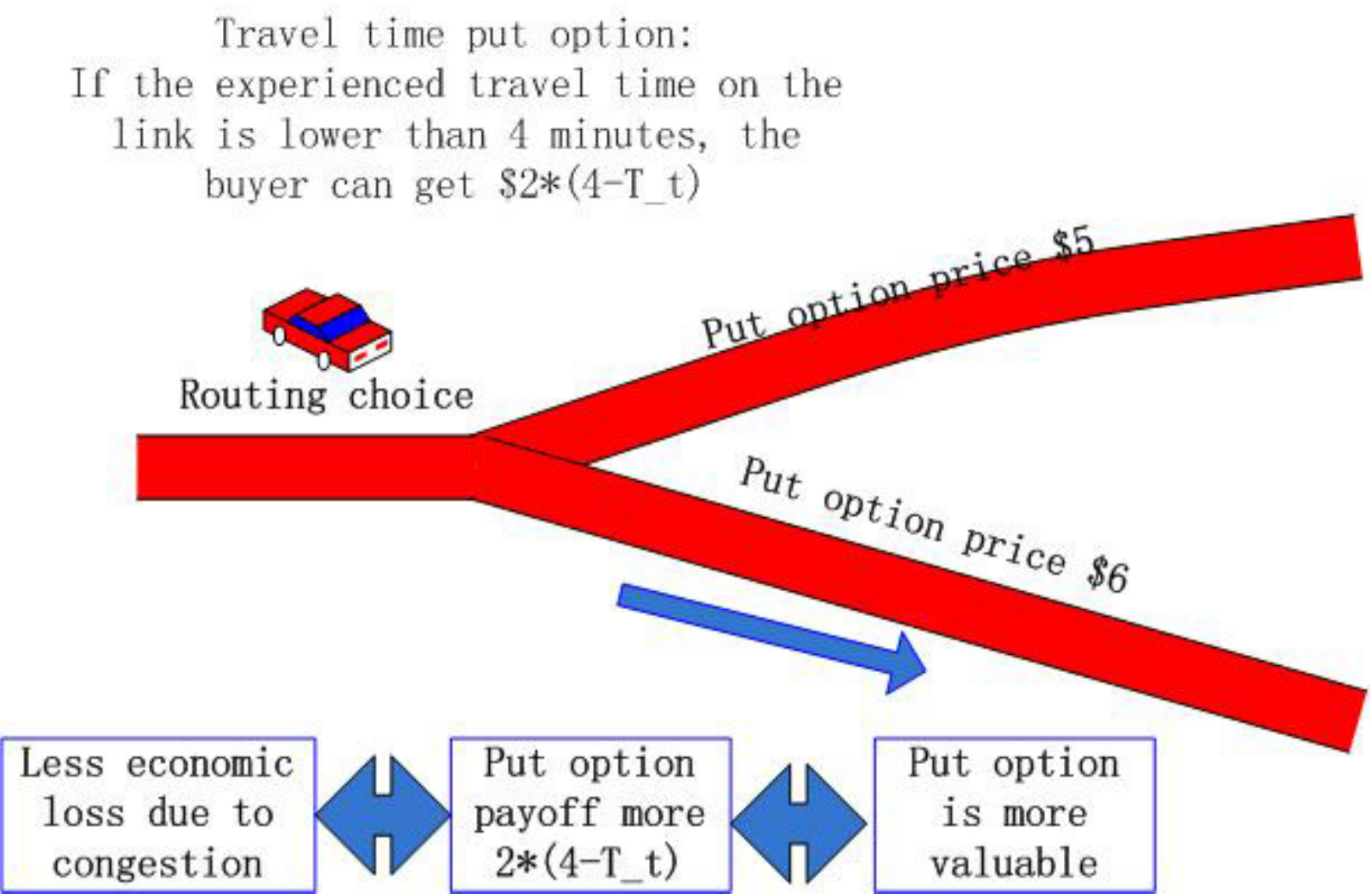}
\caption{The price of travel time derivatives indicates short term
profit and loss due to traffic conditions: Between two alternative
routing choices, a higher derivative price predicts a potentially
lower loss and hence the corresponding link is chosen by the
traveler.}\label{fig:7PredictionEffect}
\end{center}
\end{figure}

\item The second type of travel time derivative can be derived based on
the cumulative travel time measures for a long time period in the
future. For some products of this type, the economic loss due to
congestion is less in the coming year if and only if the price is
higher. Therefore, the price of travel time derivatives indicates
economic loss associated with long-term traffic conditions and a
traveler can then plan to use alternative paths or use public
transportation if derivative prices are generally low. When
considering a long-term transportation plan, travelers can check the
price of such derivatives. This category of travel time derivatives
is demonstrated in Figure \ref{fig:7PredictLongterm} and the CDD
index option cited in Table \ref{tab:7-ttoption} belongs to this
category.

\begin{figure}[ht]
\begin{center}
\includegraphics[angle=0, width=0.7\textwidth]{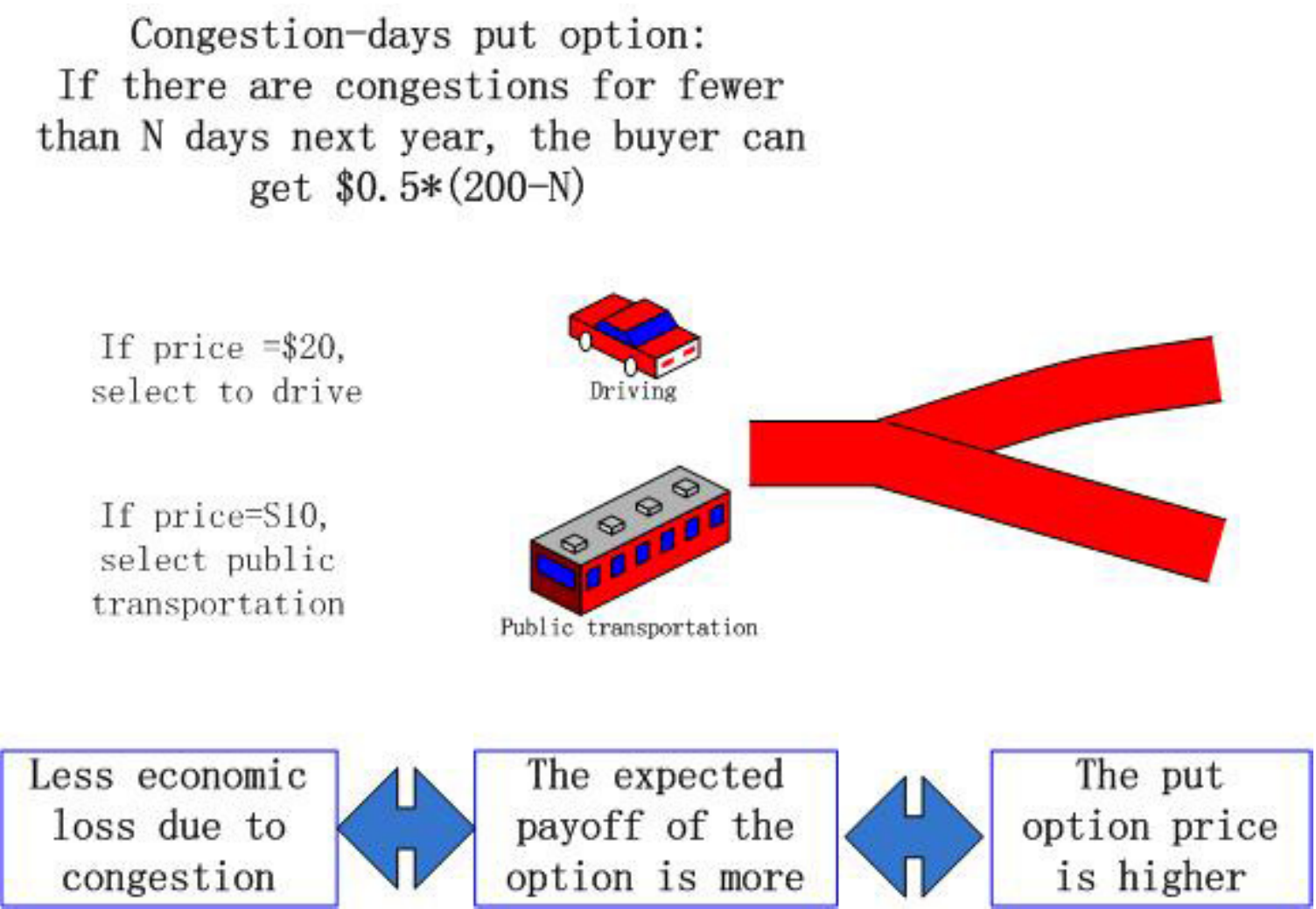}
\caption{The price of travel time derivatives indicates profit and
loss due to long term traffic conditions: Between two alternative
travel plans, a higher derivative price predicts a potentially
overall lower loss in a year and hence the traveler chooses to
drive.}\label{fig:7PredictLongterm}
\end{center}
\end{figure}

\end{enumerate}

%In summary, as flexible payoff functions can be defined, travel time
%derivatives can provide a payoff according to travelers' experienced
%travel time in future to reduce potential economic costs due to
%traffic delays, which is more advantageous over traditional
%insurance.

In this way, travelers' tolls are based on expected future traffic
conditions. If future traffic conditions are expected to be good
(bad), the potential future loss is less (more), the payoff is less,
and the toll will be less (more). Therefore, travelers could
consider taking alternate routes to avoid congestion and reduce
travel costs. This flexibility in payment makes the prices of travel
time derivatives effective predictors of future travel times.
Travelers can forecast the travel time of a path by researching the
prices of the corresponding path. This would change traveler
behaviors and help them to reduce real-time costs due to traffic
delays.

%\subsection{Travel time derivatives can provide protections to
%firms or businesses whose profit is related to traffic conditions}

\subsection{Travel time derivatives can diversify risk for financial markets
}

Travel time derivatives can provide new underlings into the
investment universe, whose change is relatively independent from
existing financial assets. In portfolio theory, the diversification
of investments into different asset classes is a recommended
practice. For a given set of investments, the lower the correlation
between assets, the less the total risk, as in
\cite{luenberger1998investment}. Most traditional asset classes
(equity or bond) are derived from the capital of companies and thus
are highly correlated in nature. The correlations between travel
time and equity/bond classes are lower than the correlations among
different equities/bonds, and the low correlation serves to
diversify the portfolio. As investors recognize travel time
derivatives as effective risk-reducing elements in their portfolios,
they will invest money in the market.

%The basic risk diversification between the financial system and the
%traffic system is shown in Figure ~\ref{fig:risktransfer}.
%\begin{figure}[htm]
%\begin{center}
%\includegraphics[angle=0, width=0.6 \textwidth]{figures/TD_ARC1}
%\caption{Alternative risk transfer between the transportation and
%financial industries} \label{fig:risktransfer}
%\end{center}
%\end{figure}
%%

Travel time derivatives have the potential to hedge risks in other markets for derivatives on non-tradable underlyings or commodities. For instance, since weather conditions are correlated with travel time, travel time derivatives could serve as a useful hedging tool for investors who invest in the weather derivative market. Similarly, energy consumption and CO2 emission levels have been traded in the market, and because their performance is highly correlated with that of the traffic system, travel time derivatives could be a useful tool to hedge against risk for investors in these markets.

Moving beyond the transportation sector, during the 2020-2022 period, both traditional financial markets and the crypto market experienced severe stress due to COVID-19 and interest rate hikes implemented to counter high inflation \cite{cornelli2023crypto} \cite{GRUENBERG2023bankingstress}. A new category of derivatives linked to the economy, but not fully subject to the leverage of cryptocurrency or the credit quality of traditional banking, could be a promising hedging tool to help investors navigate future market stress.

\section{Potential participants and market making}

%Following the necessity analysis above, potential participants in travel time derivative markets are introduced, and critical market-making factors are addressed.

In light of the necessity analysis outlined earlier, it is important to consider potential participants in travel time derivative markets and the key market-making factors that will need to be addressed. This includes identifying various parties who could benefit from travel time derivatives, such as transportation service providers, government agencies, insurance companies, and institutional investors. Additionally, critical market-making factors that will need to be addressed include designing contracts that are attractive to buyers and sellers, ensuring sufficient liquidity, and developing appropriate pricing models to accurately reflect the risks associated with travel time derivatives. Addressing these factors will be vital to the success of travel time derivative markets.

To provide the hedging effect, there are two types of travel time derivatives for the two sides of the market. Type B: When the specified travel
time is expected to be high, a leveraged reward is available to the
buyer. Type H: When the specified travel time is expected to be low,
a leveraged reward is available to the buyer. Accordingly,
participants with different risk profiles will buy different travel
time derivatives to hedge their risk; buyers of Type B are the
participants who benefit (lose) from good (bad) traffic
conditions; buyers of Type H are the participants who lose (who
benefit) from good (bad) traffic conditions. The potential
participants and their roles are summarized in Table
~\ref{tab:Depaticipants}.

\begin{table}[ht]
\caption{Different participants (B means benefit in good quality of
service(QOS); H means hurt in good QOS; market makers and investors
can hold both types of derivatives.)} \label{tab:Depaticipants}
\centering
\begin{tabular}{|c|c|c|}
\hline
Parameter &Hedging Motivation & Type \\
\hline
Individual travelers & Traffic delay, business delay & B \\
 & extra charges due to bad QOS& \\
\hline
 Cargo transportation& Traffic delay due to bad QOS & B\\
 &QOS&\\
\hline
Tourism industry & Traffic delay due to bad QOS & B\\
\hline
Event organizers& Traffic delay due to bad QOS&B\\
\hline
Municipal management&Traffic delay due to bad QOS& B\\
\hline
Insurance companies&Loss due to vehicle accidents due to bad QOS&B\\
&Repackaging of vehicle insurance&\\
\hline
Gas company&Low overall gas consumption&H\\
\hline
Owners of Toll roads& Low profit due to good &H\\
&QOS on toll free roads&\\
\hline
Vehicle maintenance&Fewer accidents and business loss due to good QOS &H\\
\hline
 Auto companies&Fewer needs for new autos due to good QOS&H\\
\hline
Public transportation&Less business due to good QOS&H\\
\hline
 Taxi companies&Less business due to good QOS&H\\
\hline
Alternative transportation (train) &Less business due to good QOS&H\\
\hline
 Banks&Market Making&B/H\\
 \hline
 Traffic detection agencies&Measurement providers &B/H\\
 GPS companies&&\\
 \hline
 Portfolio managers&Risk diversification&B/H\\
 &Speculation&\\
 \hline
 Project management&Project financing &B/H\\
 &Speculation&\\
 \hline
\end{tabular}
\end{table}

As a newly introduced market, the market making of travel time
derivatives is critical and challenging, \cite{panayides2004role}.
Market makers match buyers with sellers to enable smooth trading
activities; they also provide liquidity to the market by holding
short term positions; due to their efforts, the market price of
financial derivatives is determined and maintained. To make a
profit, market makers quote both a buy and a sell price in
derivative contracts, which differ on the bid-offer spread, and they
use hedging strategies to control their risk. Investment banks are
typical market makers for financial derivatives. With appropriate
pricing methods and suitable trading exercise, the total profit for
investment banks is positive, which motivates them to operate the
business, following the general mechanism in the current financial
derivative markets. On the other hand, the counter parties to the
market parties will seek protection from the market and their total
profit are negative, which can be interpreted as the cost that they
pay to hedge risk due to travel time uncertainties in the future.
Important factors that should be considered for market making
include the following:

 \begin{enumerate}

\item Market micro structure will be crucial in determining the operation
of the market. There are numerous links/paths in transportation
networks, and a large number of derivative contracts can be written
based on their experienced travel time. Conversely, when this new
market begins operating, the trading activity will be low.
Therefore, the market may encounter liquidity issues, where smaller
trading amounts can drive the prices, and thereby increase price
volatility. Several measures can be taken to minimize potential
liquidity issues, including restricting the number of products on
the market, building temporary liquidity reserves, and so forth.
Related discussions for other types of derivatives can be found in
\cite{mackenzie2003constructing},
 \cite{dubil2007economic}, \cite{wolfers2006five} and
\cite{zitzewitz2006price} .

\item Market scale is important for the sustainability
 of the derivatives markets. A survey conducted by the U.S. Department of Commerce in 2004 estimated that
 approximately 30\% of the total U.S. GDP is exposed to some degree of weather risk, \cite{Finnegan2005}.
 This considerable percentage leads to the necessary liquidity and prosperity of a weather derivative market.
 In the transportation industry, the percentage needs to be estimated and a larger percentage means
 more potential market participants. There is significant amount of research on the cost of travel time,
 which can roughly be measured in the annual revenue raised by road tolls.
 For example in \cite{sugiyanto11estimation}, it is stated "there were \$63.2
 billion in actual congestion costs in the 85 urban areas [in the U.S.] in 2002.
  It is estimated that public transportation saved an additional
  \$20 billion in congestion costs for this group."
  The billion-dollar congestion costs imply
  a significant impact of traffic delays
   to individual travelers, which motivates them to hedge their risks.
   More profoundly, the companies, the profits of which are changed by
   traffic service, may have more freedom to purchase and trade travel time derivatives,
   which should be further estimated. The sum of all related profit and costs add to
    the potential for travel time derivative markets, \cite{harford2006congestion} and \cite{nash2001pricing}. 

    More recently, According to \cite{energegov2022}, Transportation plays an important role in the U.S. economy and contributed 8\% to U.S. Gross Domestic Product (GDP) in 2020. Transportation is the fourth-largest contributor (behind housing, healthcare, and food) to the national GDP, which measures the monetary value of goods and services in the United States.
    Per \cite{btsgov2022}, In 2021, the demand for transportation (\$1.9 trillion) accounted for 8.4 percent of GDP. Please see Figure \ref{fig:gdp-transportation}. While the figures shows GDP directly linked to transportation, there are even more implicit impact of transportation to GDP because the delivery of raw materials and products of a great many other industries rely on quality of service of the transportation system.

\begin{figure}[htb]
\begin{center}
\subfloat[Contribution to U.S GDP by Category in 2020]% \quad on the next line adds spacing
{\includegraphics[angle =0,width=0.45
\textwidth]{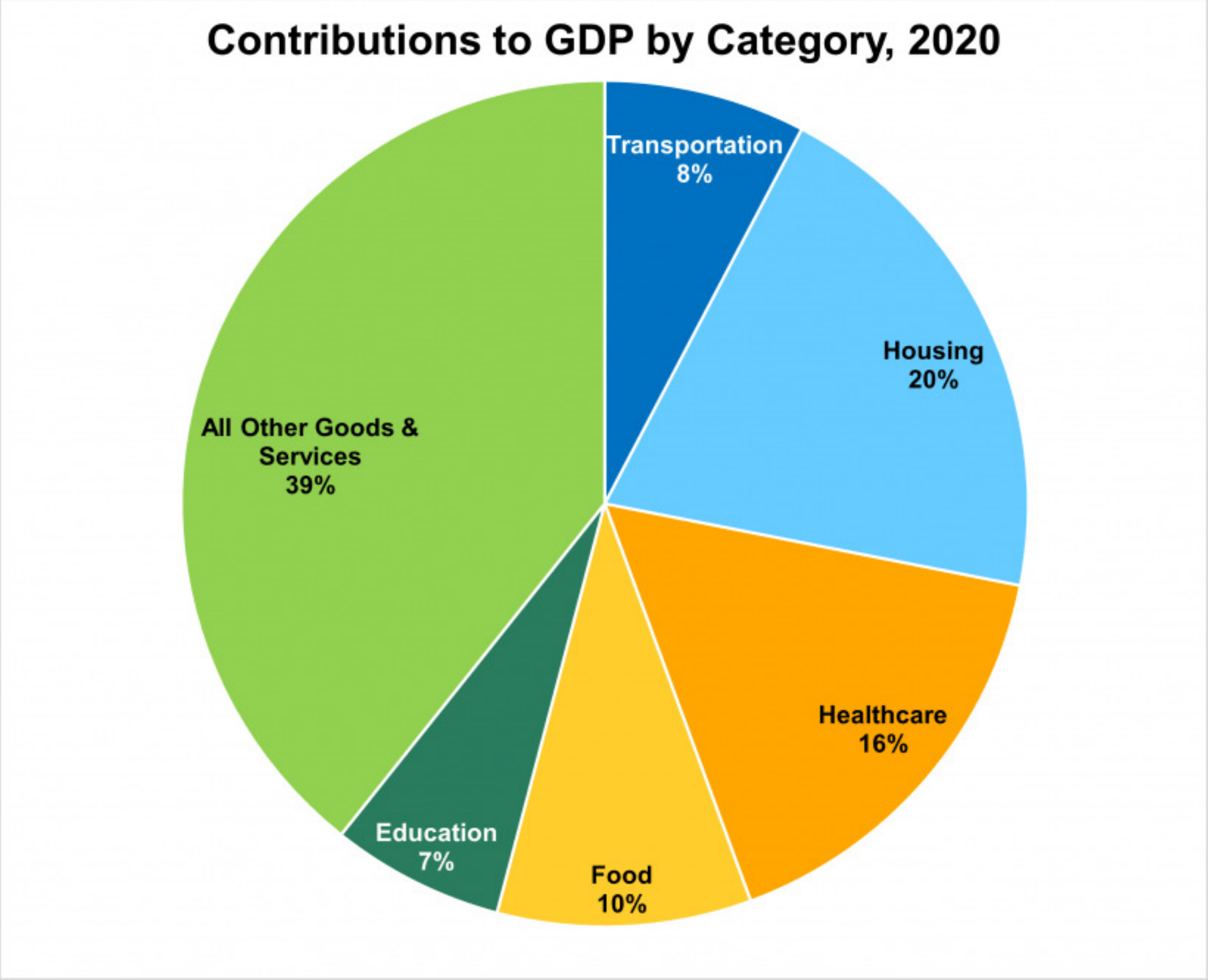}}\quad
\bigskip
\bigskip
\subfloat[Contribution of Transportation to U.S GDP over years] {\includegraphics[angle=0, width=0.45
\textwidth]{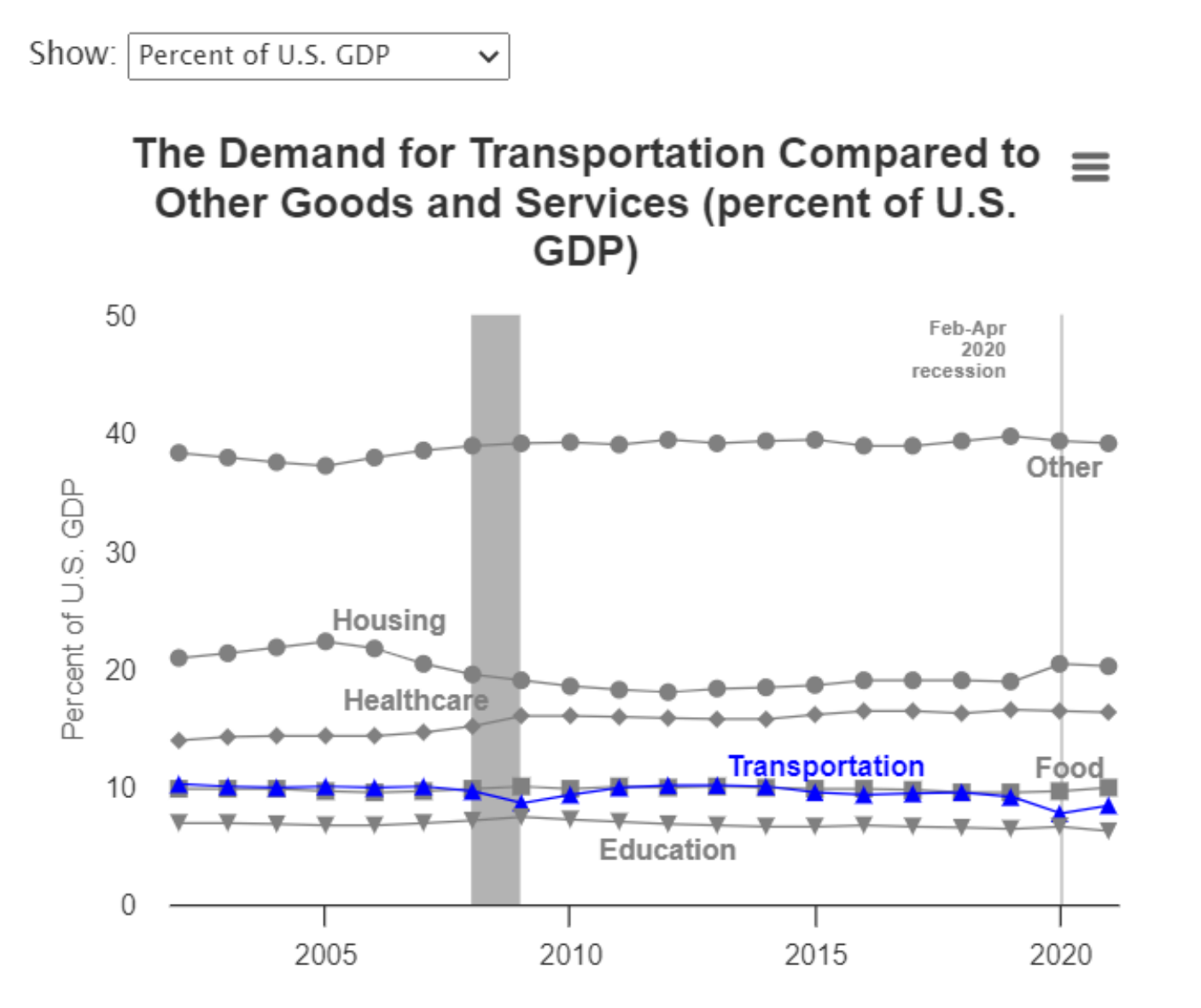}}\quad
\bigskip
\bigskip
\caption{Contribution of Transportation to U.S GDP}
\label{fig:gdp-transportation}
\end{center}
\end{figure}

\item
A healthy market for travel time derivatives also requires
appropriate legal regulations. As observed in traditional financial
markets, malicious insider trading or market manipulation can occur
if the participants know additional information, which may change
future travel time through illegal sources. Moreover, unlike
traditional underling assets, travel time is the aggregate effect of
traveler behavior nearby, so the potential of travel time
derivatives in changing traveler behavior may lead to possibility of
manipulating future travel times and hence price of travel time
derivatives through intentional routing guidance. To prevent such
un- desired cases and regulate the travel time derivatives market,
appropriate policies or laws should be issued.

\end{enumerate}

Based on the settings above, a market can potentially be established
for travel time derivatives. The major products for this potential
market are presented in the following section.

\section{Design of Travel Time Derivatives}

In general, the underlying asset of travel time derivatives is some
measure of future travel times. Investors receive cash flows in
proportion to travel time related measures as their payoff and in
order to purchase such derivatives, investors have to pay a price.
Based on the introduction to travel time derivatives in previous
sections, several classifications can be applied to travel time
derivatives:

\begin{enumerate}
\item By contract type, travel time derivatives can be classified into
futures, options, etc;

\item By measurement places used in the derivative, they may be classified
as derivatives based on one path or several paths, which is hence
based on an index of travel time;

\item By the time span of the underlying travel time measures, travel time
derivatives can be classified as instantaneous or long term based;
\item By whether the buyer gets a payoff when traffic is good
or bad, travel time derivatives can be classified as Beneficial (B)
versus Hurting (H).

\end{enumerate}
This section continues to introduce more mathematics for describing
the travel time derivatives.

\subsection{Standard travel time measurements}

In order to define products based on travel time measures, a
standard measurement of travel time must be defined.

\begin{thm}
A standard measurement of travel time on a specific path and time is
the average travel time reported from specific travel time data
providers on that path within s small time interval around that
time.
\end{thm}

\begin{thm}
A standard measurement plan of travel time on a specific path on a
specific day is a set of standard measurements which are collected
at a pre-defined time of day. The daily mean of a standard
measurement plan is the mean value of such measurements.
\end{thm}

In the above definitions of travel time derivatives market products,
all travel time values for a given time in a day are based on a
standard measurement, and all travel time values for a given day are
based on the standard measurement plan. Each observation is selected
by specifying the arrival time to the path. To provide adequate
measurements of travel time to support the trading and pricing of
travel time derivatives, a loop detector is recommended. The reasons
for this choice include:

\begin{enumerate}
\item Pricing of travel time derivatives should be based on periodical
travel time measurements so that classical stochastic analysis can
be used to model travel time and corresponding models can be
calibrated. As is summarized in Section 2.1, site-based measurement
such as loop detectors can yield such periodical estimations based
on occupancy and flow, and hence data from them are suitable for the
study of travel time derivatives.

\item Pricing of travel time derivatives should be based on average
travel time on the given path to prevent individual measurement
error from introducing instability in derivative prices. Loop
detectors yield estimation of average travel time based on occupancy
and flow, which satisfies this requirement and relieves related
concerns.

\end{enumerate}

\subsection{Standard spatial travel time index and equivalent return rate}

First, a spatial travel time index should be designed. This index
can be a weighted average of the latest travel time in downtown
areas of major cities in the U.S. A national travel time index is an
objective reference for trading and a good symbol for the
transportation industry. The definition is given below, and local
travel time indexes can be designed in a similar fashion.

\begin{thm}
Spatial Travel Time Index
$$T_{us}=\sum T_i*\alpha_i$$
where $T_i$ is the travel time in selected places within a given
area.
\end{thm}
%\newnot{symbol:Tus}

For example, a Spatial Travel Time Index could be constructed as the
weighted average of the realtime travel time in downtown New York (a
section of Fifth Avenue), downtown Chicago (a section of Michigan
Avenue), downtown Los Angeles (a section of Sunset Boulevard), and
downtown Houston (a section of Main Street). This index can be
viewed as an average traffic index on the quality of service of the
urban transportation system in the United States, which shows the
national service level of urban traffic systems. The return of this
index, volatility, and its sharpe ratio can then be used as
references when pricing travel time derivatives. Note the Sharp
ratio is the ratio between access return and volatility of this
derivative, and excess return is the extra return of this index
relative to the risk-free bond.

Travel time indexes can also be defined for a given area, if taking
the average travel time in the major avenues of Manhattan can be
used to indicate the general traffic conditions on Manhattan island.
Travelers can trade over such indexes to compensate for the waste of
time and economic loss due to traffic delays.

To provide a more practical example, a U.S. Congestion Day index
futures contract can be defined in a similar fashion as the Canadian
Degree Days index HDD futures contract displayed in Table
\ref{tab:7-ttoption}. In this contract, the average daily travel
time (which can be measured using certain sampling schemes) minus a
predefined travel time value floored at zero is summed over a given
calendar month for several urban transportation routes in the U.S.
Suppose the buyer purchases the contract at price P; initially, if
travel times on the specified urban routes are generally higher and
this sum is potentially higher in a given month, the price of the
contract would increase. In this case, the buyer profits from the
price increase and increased payoff G-P to compensate potential
losses due to high travel time. This contract can be designed and
traded in a similar fashion as that for weather derivatives. Note
that related mechanisms are discussed in more detail in the
subsequent sections, beginning with more basic products.

\begin{table}[ht]
\caption{US Congestion Days Index (CDD) Futures}
\label{tab:7-ttoption} \centering
\begin{tabular}{|c|c|}
\hline
 Contract Size & US \$20 times the respective \\
 & CME USA Congestion Days Index\\
\hline
 Index Product Description & Congestion Degree Days (CDD) \\& for U.S. Cities\\
\hline Measurement definition&The travel time in a particular city
is reported\\& based on a specific
 route:\\
 &along Fifth Avenue between 59th st\\& and Washington Square, New York \\
&along Michigan Avenue between South Lake Drive \\& and Roosevelt Road,Chicago\\
&along Sunset Boulevard between Prospect Ave\\& and Harbar Highway, Los Angeles\\
& along Main Street between Bissonnet St\\& and Commence St, Houston \\
 \hline
 Pricing Unit & US Dollars (US \$) per index point\\
\hline
 Tick Size & 1 index point \\
 (minimum fluctuation) &(= US \$ 20 per contract) \\
\hline
Trading Hours & CME Globex (Electronic Platform) \\
(All times listed are Central Time)&SUN 5:00 p.m. - FRI 3:15 p.m. \\
&Daily trading halts 3:15 p.m. - 5:00 p.m.\\
\hline
 Last Trade Date/Time &
Fifth Exchange business day\\& after the futures
contract month, 9:00 a.m.\\
\hline
% Contract Months & HDD: Nov, Dec, Jan, Feb, Mar\\& plus Oct and Apr\\
%\hline
%Settlement Procedure & Daily Settlement Procedures for Monthly HDD Futures\\
% & Final Settlement Procedures for Monthly HDD Futures \\
% \hline
 Position Limits & All months combined: 10,000 contracts\\& See
CME Rule
42102.D.\\
%\hline
%Ticker Symbol & Calgary = A2 Edmonton = A4 Montreal = A5\\
%& Toronto=A7 Vancouver = A8 Winnipeg = A9\\
\hline
Exchange Rule & These contracts are listed with, \\
& and subject to, the rules and regulations of CME.\\
\hline
%\newnot{symbol:G}
%\newnot{symbol:Pprice}
\end{tabular}
\end{table}

\subsection{Design of derivative products on travel time}

\textbf{Type 1 Basic options for a specific link at a given future
time point}

The simple derivative based on the experienced travel time at a
future time is defined as follows, and an example is given
afterwards.
\begin{thm}

Call option on a certain travel time. Consider the link $l$ a
specific time instant $t$ in the future. If the travel time shown by
the standard measurement at $t$(denoted as $T$) is higher than a
given $K$, then there is a payment $\alpha(T-K)$ to the option
buyer; if lower, there is no payment.$\alpha$ is the leverage
coefficient.
\end{thm}
\begin{thm}

Put option on a certain travel time. Consider the link $l$ a
specific time instant $t$ in the future. If the travel time shown by
the standard measurement at $t$(denoted as $T$) is lower than a
given $K$, then a payment $\alpha(K-T)$ is available to the option
buyer ; if higher, there is no payment.$\alpha$ is the leverage
coefficient.
\end{thm}

Consider Broadway in New York City from 20th to 60th Streets. If the
travel time shown by its standard measurement entering at 10 a.m. on
January 1, 2011, is equal to 70 minutes and the threshold value is
set as 60, then there is a leveraged cash back to the buyer \$ $10*
* (70 - 60)$; if the experienced travel time is lower than 60
minutes, the buyer gets nothing.

\textbf{Type 2 Futures on congestion-days}

After establishing basic options for a specific link at a given
future time point, the futures written on the high congestion days
(HCD) and low congestion days (LCD) are then designed. As a basic
concept, the definitions of HCD and LCD are given below:

\begin{thm} High Congestion Days(HCD) and Low Congestion
Days(LCD) in discrete time settings

 Let $T_i$ denotes the mean of a standard measurement plan on day $d_i$ and $C$ as a specified reference value. The high congestion-days, $HCD_i$, and the lower congestion-days, $LCD_i$, on that day are defined as $HCD_i=max(T_i-C,0)$ and $LCD_i= max(C-T_i,0)$ respectively. In other words, $HCD$ is the extra amount of travel time spent on that day compared to the reference value $C$, and $LCD_i$ is the amount of travel time savings compared to the reference value $C$.

Then the HCD for a given time period $[t_1,t_2]$ is defined as the
sum of the HCD on all the days in that period, given a fixed number
of measurements.
 $$HCD(t_1,t_2)=\sum_{i=1}^{n} HCD_i 1_{d_i\in [t_1,t_2]}$$

The LCD for a given time period $[t_1,t_2]$ is defined as the sum of
the LCD on all the days in that period, given a fixed number of
measurements.
 $$LCD(t_1,t_2)=\sum_{i=1}^{n} LCD_i 1_{d_i\in [t_1,t_2]}$$, as the HCD/LCD for the time period.
\end{thm}

In a continuous setting, the payoff functions should be defined as
follows:
\begin{thm} High Congestion Days(HCD) and Low Congestion
Days(LCD) in continuous time

 Given a threshold $T$, the HCD for a given time period $[t_1,t_2]$ is defined as

$$HCD(t_1,t_2)=\int_{t1}^{t2}max(T_t-T,0) dt$$

the LCD for a given time period $[t_1,t_2]$ is defined as
$$LCD(t_1,t_2)=\int_{t1}^{t2}max(T-T_t,0) dt$$

\end{thm}

This pair of products shows the cumulative performance of the path
compared to some average reference. Its price will reflect market
participants' expectations of the quality of service on the path;
hence, its price can predict the long term traffic status on the
path.

 \textbf{Type 3 Congestion-days options}

Congestion days options are options based on the average performance
of a path in a future time window. The definitions are given first,
followed by an example.

\begin{thm} Call options on high congestion days. Denote $K$ as the strike value:

 The payoff of a HCD call is $$V = \alpha max(H_n-K,0)$$
The payoff of a LCD Call is $$V = \alpha max(L_n-K,0)$$
\end{thm}
\begin{thm} Put options on high congestion days. Denote $K$ as the strike value:

 The payoff of a HCD put is
$$X = \alpha max(K-H_n,0)$$
The payoff of a LCD put is
$$X = \alpha max(K-L_n,0)$$
\end{thm}

Consider Broadway in New York City from 20th to 60th Streets. If the
mean travel time on it is greater than 60 minutes, then a surplus
$T-60$ is recorded as a congestion day; otherwise 0 surplus is
recorded. Then all these surplus values are added together for one
year with 365 days. If the sum $S$ equals 2000 and so is larger than
$K = 1500$, then there is a leveraged cash back $10*(S-K)$ where 10
is the leverage ratio; if not, the buyer receives nothing. This is
an example of an HCD call option. This pair of products leverages
the buyer's gain according to long term traffic status in the
future. Compared to the futures, the options provide further
leverage, and the buyers can get more return/loss if traffic
conditions change. In buying such products, a traveler will change
travel patterns accordingly. In this sense, options on travel time
are effective in changing a traveler's behavior.

\textbf{Type 4 Futures on cumulative travel time}

This product is the futures contract written on the cumulative
travel time in a future time period. The cumulative travel time
index is defined first.

\begin{thm} Cumulative travel time
index (CTT) in discrete time settings.

 The CTT index over a time interval $[t_1,t_2]$ is defined as the sum of the daily standard
measurement plan in a given time period.

$$CTT(t_1,t_2)=\sum_{i=1}^{n}T_{t_i} I_{t_i\in [t_1,t_2]}$$.
\end{thm}

\begin{thm} Cumulative travel time index in a future time
period(CTT) in continuous time settings.

The CTT index over a time window $[t_1,t_2]$ is defined as the
integration of travel time in that time window.
$$CTT(t_1,t_2)=\int_{t1}^{t2}T_t dt$$
\end{thm}

The payoff of the futures on CTT is in direct proportion to the
travel time that a traveler experiences over a given time period. It
is an alternative measure of the long term quality of service to
HCD/LCDs.

 \textbf{Type 5 Options on cumulative travel time }
 These products are the options written on the CTT index. Their payoff functions are given as follows:

\begin{thm} Call options on cumulative travel time. Denote $K$ as the strike value:
 The payoff of a HCD call is
$$V = \alpha max(CTT-K,0)$$
\end{thm}\bigskip
\begin{thm} Put options on high congestion days. Denote $K$ as the strike value:
 The payoff of a HCD put is
$$X = \alpha max(K-CTT,0)$$
\end{thm}
Again, the options on CTT provide greater leverage than other forms
of road tolls; therefore, they can potentially change a traveler's
behavior more effectively.

\section{Pricing derivatives on travel time}

To price travel time derivatives, the underlying travel time series
is first selected as a continuous time mean reverting process with
trend and seasonality adjustments.
%
%As a model selection process, There are different advantages for
%different models, and the data in different links may show different
%characteristics; a family of alternative models is given below.
 Due to variation in traffic conditions across different links, the travel times on different links
 may be fitted to models with different orders. To provide such flexibility, a family of alternative models
are introduced in this section. Model selection is conducted based
on empirical data according to statistical principles and pricing
methods are discussed.

\subsection{Alternative stochastic processes for modeling travel time}

This section provides an overview of relevant stochastic processes used for modeling time series data similar to road travel time and pricing financial derivatives based on non-tradable underlyings. Mean reverting processes are frequently used in financial literature to model time series that tend to move back to their average over time, such as interest rates and weather derivatives \cite{hull1990pricing}. For example, Benth and Benth (2007) \cite{benth2007putting} analyzed weather derivatives traded at the Chicago Mercantile Exchange and modeled temperature dynamics as a continuous-time autoregressive process, allowing for pricing of futures and options on weather. In Gyamerah (2019) \cite{gyamerah2019hedging}, a machine learning model was used to identify drivers of maize yield, and a mean-reverting model was proposed with a time-varying speed of mean reversion, seasonal mean, and local volatility that depended on local average temperature. Alfons (2022) \cite{alfonsi2022stochastic} developed a stochastic volatility model for average daily temperature, which was calibrated using daily data from eight major European cities, and pricing was done using Monte-Carlo and Fourier transform techniques.

Alternatively, local risk minimization indifference pricing is also a promising methodology for pricing derivatives under incomplete markets. Rene (2013) \cite{aid2013structural} used the local risk minimization approach to price and hedge energy derivatives in incomplete markets by specifying the set of infinitely many equivalent martingale measures and identifying the minimum measure with zero risk premium. Chen (2021) \cite{chen2021indifference} studied utility indifference pricing of derivatives based on untradable assets in incomplete markets using a symmetric asymptotic hyperbolic absolute risk aversion (SAHARA) utility function.

To model travel time, similar stochastic processes can be used, and specific patterns of travel time are considered in the fitting process. Ke (2009)\cite{wan2009travel} introduced the idea of financial derivatives based on travel time, and further work on the properties and pricing methodology was summarized in Ke (2014)\cite{wan2014estimation}. Theoretical foundations for discrete ARMA models are summarized in detail in Janqing(2017) \cite{fan2017elements}, and background on continuous ARMA processes is provided by Brockwell (2001) \cite{brock2001continuous} and Brockwell (2010) \cite{brockwell2010carma}.

\subsubsection{Continuous autoregressive moving average (CARMA) process }

Let $(\Omega ,\mathcal{F}, {\mathcal{F}}_{\{t>0\}}, P)$ be a
complete filtration probability space. A random variable is a
mapping $X:\Omega\rightarrow R^d$, if it is $F$-measurable, whereas
a family of random variables depending on time t, ${X_t}_{t>0}$ is
said to be a stochastic process. A process $X_t$ is $F$-adapted if
every $X_t$ is measurable with respect to the $\sigma$-algebra $F$.

Then a continuous-time Gaussian autoregressive and moving
average process (CARMA) can be used to fit the travel time process
The process $X_t$ in previous section is the simplest CARMA($1,0$)
process. By definition, a CARMA process $Y_t $ is defined
symbolically to be a stationary solution of the stochastic
differential equation:

$$a(D)C_t = b(D)B_t$$

with coefficients $a_1,a_2,\cdots,a_p$ and $b_0,b_1,\cdots,b_q$ for
$p > q$

$a(z)= a_0z^p + a_1 z^{p-1}+\cdots+a_p$

$b(z)=b_0+b_1z+\cdots+b_qz^q$

The operator $D$ denotes differentiation with respect to $t$, which
is in the formal sense for the Brownian Motion
\cite{brock2001continuous}. Due to the fact the derivative of $B_t$
does not exist with probability 1, the process is represented
further in the following state space representation

$$C_t=BX_t$$

and

$$dX_t=AX_tdt+edB_t$$

 with

$B=[b_0,b_1,\ldots,b_q,0,\ldots,0]$

$X_t=(X_{t,0}; \ldots ;X_{t,p-1})$ and the first $p-1$ element of
$X_t$ is defined as

$$X_{t,j}-X_{0,j}=\int_0^t X_{u,j+1}du$$

$A$=\[ \left[ \begin{array}{ccccc}
0 & 1 & 0 & \cdots &0 \\
0 & 0 & 1 & \cdots &0\\
\vdots & \vdots & \vdots & \ddots & \vdots\\
0 & 0 & 0 & \cdots & 1\\
-a_p & -a_{p-1} & -a_{p-2} & \cdots & a_1  \end{array} \right]\]

 $\textbf{e}=[0,0,0,0,0,0,1]^T$

when $p=1$, $A$ is defined as $-a_1$

By applying the multidimensional Ito Formula, the solution to the
S.D.E above is below:

$$X_t=e^{At}X_0+\int_0^t e^{A(t-u)}\textbf{e}dB_u$$

The process above is stationary and well defined when $p > q$. If $p
\leq q$, the covariance function does not exist and the spectral
density does not exist. However, the process is further defined as a
general random process (GRP) by Brockwell (2010) \cite{brockwell2010carma} for the
study of the derivatives of CARMA($p,q$) processes. In the following
sections, the paper focuses on the continuous time derivative
pricing based on CARMA($p,q$) process for the $p > q$ case, while it
is recommended that Monte Carlo simulation be conducted based on the
fitted discrete time models to approximate the derivative prices for
if $p \leq q$.
%
%if $p=0$ $$X=\sum_{j=0}^q b_j B^{j+1} $$
%%
%if $p>0$ $$X=\sum_{j=0}^q b_j X^{j} $$
%
%where
%
% $B^{(1)}=\int_R dB_t$ is the first order integration of Brownian Motion
%
% $B^{(j+1)}=\int_R d B^{(j)}$ is the $j+1$-th order integration of Brownian Motion
%
% $X^{(j)}=e_1^{'}A^jB^{(1)}*g+\sum_{k=p-1}^{j-1}B^{(j-k)}e_1^{'}A^ke_p$
% is the solution to the process in the GRP sense.
%
% $B^{(1)}*g(1)=\int_R g(-t) B_t$ is the convolution of $g$ and $B_t$
%
% $g(t)$ is the kernel function which is defined in
% \cite{brockwell2010carma}.

%The definition when $p \leq q$ exists in the general random process
%sense. Due to the special treatment, the model corresponding to $p
%\leq q$ case is not used for pricing travel time derivative while
%Monte Carlo simulation is recommended based on the fitted discrete
%ARIMA model.

\subsection{Modeling travel time processes}

In this subsection, empirical data are used to select the model to
describe travel time processes, and corresponding parameters are
estimated. The driving process is identified as an ARIMA model, and
corresponding continuous versions are given. The model is then used
to price derivatives in later sections. First, the mean reverting
model is re-parameterized as follows:

\begin{eqnarray}
T_t = r_t+ s_t + Y_t\mbox{, } t = 0, 1, 2, \ldots
\end{eqnarray}
 where $r_t$ is the trend part, $s_t$ is the season part, and $Y_t$ is the driving process that shapes the noise term.
 Define $c_t=r_t+s_t$ as the sum of the trend and seasonal parts. The terms are explained separately as follows:

\begin{enumerate}
\item
The trend part is a linear function over time
$$r_t=a+bt$$
\item
Seasonal parts are as follows:
$$s_t=b_t+w_t$$
\begin{enumerate}
\item The daily part is:
$$b_t=k+\sum_1^{I_s} a_t(sin2i\pi(t-f_i)/T_d+\sum_1^{J_s} b_t(cos 2i\pi(t-g_i)/T_d)$$
%$$\sigma^2_t=c+\sum_{i=1}^{I_\sigma}c_i sin(2i\pi t/24)+\sum_{i=1}^{J_\sigma}d_i cos(2j\pi t/24)$$
%

\item The weekly part is:

$$w_t=k+\sum_1^{I_s} a_t(sin2i\pi(t-f_i)/T_w+\sum_1^{J_s} b_t(cos 2i\pi(t-g_i)/T_w)$$

\item Another alternative is the 10-parameter model given
in \cite{schrade2003}:
$$V(t)=\mu+\sum_{i=1}{3}k_i \phi(t,\mu_i,\sigma_i)$$
 However, the trigonometric functions are selected as the basis function, because they are orthogonal and suitable to describe the
 periodical pattern of travel time.
\end{enumerate}
\item $Y_t$ is the driving process. It is a stochastic process
%that may contain
%a mean reverting part and a simpler stochastic part, which is
%usually a Brownian motion.
and the type of process is determined based on empirical data. For
the data in this section, it is defined that $C_t=dY_t$ and $C_t$ is
modeled as a CARMA($p,q$) process.

\end{enumerate}

The data set is the travel time data set from California PEMS. Data
were collected in November 2012, on the 80-E path between
80-E/Cummings Skwy and 80-E/Maritime Academy, as presented in Figure
\ref{fig:5-network}. The travel time data are gathered from loops
captured every 5 minutes. The calibration process is as follows:

\begin{figure}[htb]
\begin{center}
\subfloat[The path]% \quad on the next line adds spacing
{\includegraphics[angle =90,width=0.35
\textwidth]{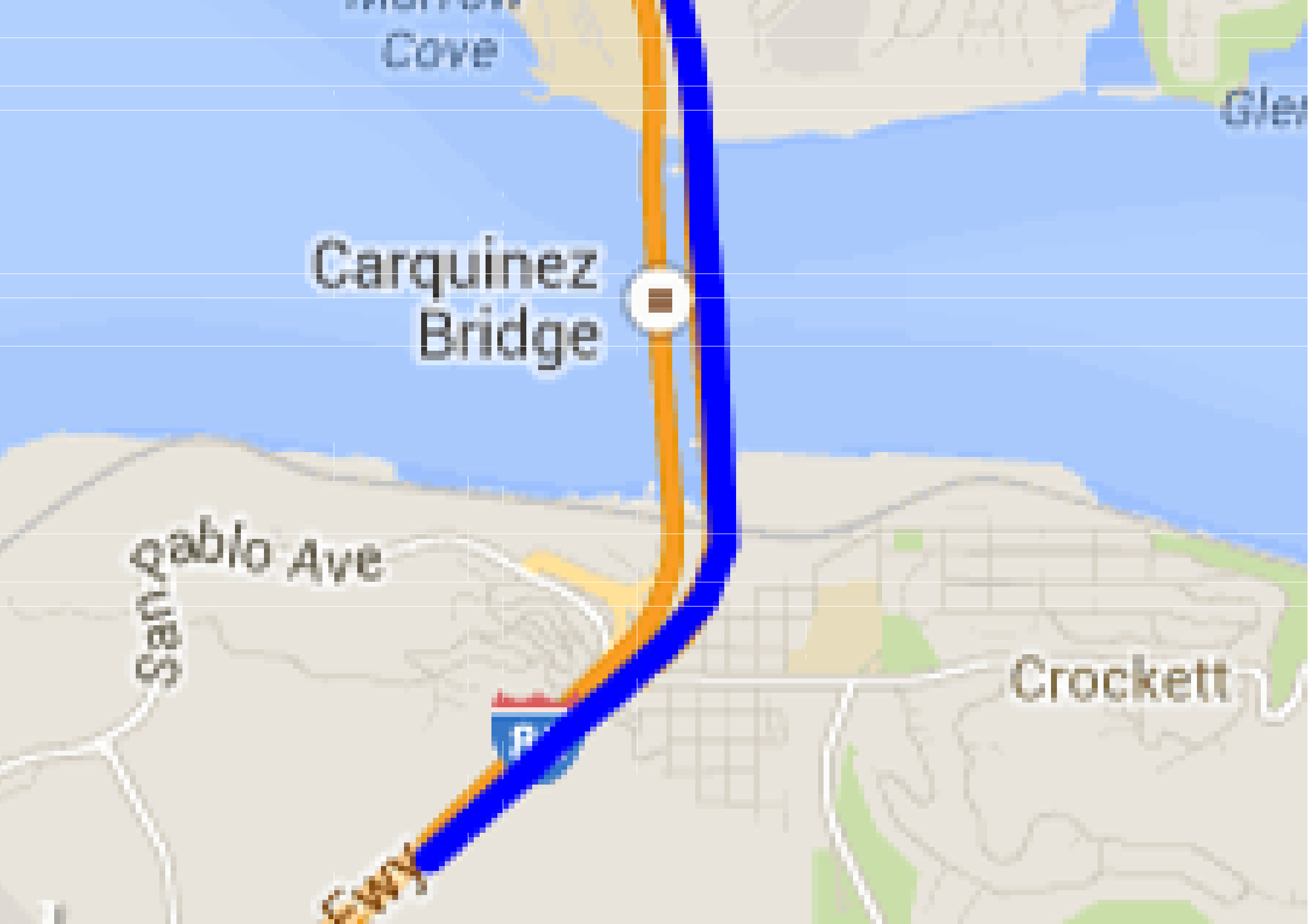}}\quad
\bigskip
\bigskip
\subfloat[The original data] {\includegraphics[angle=0, width=0.45
\textwidth]{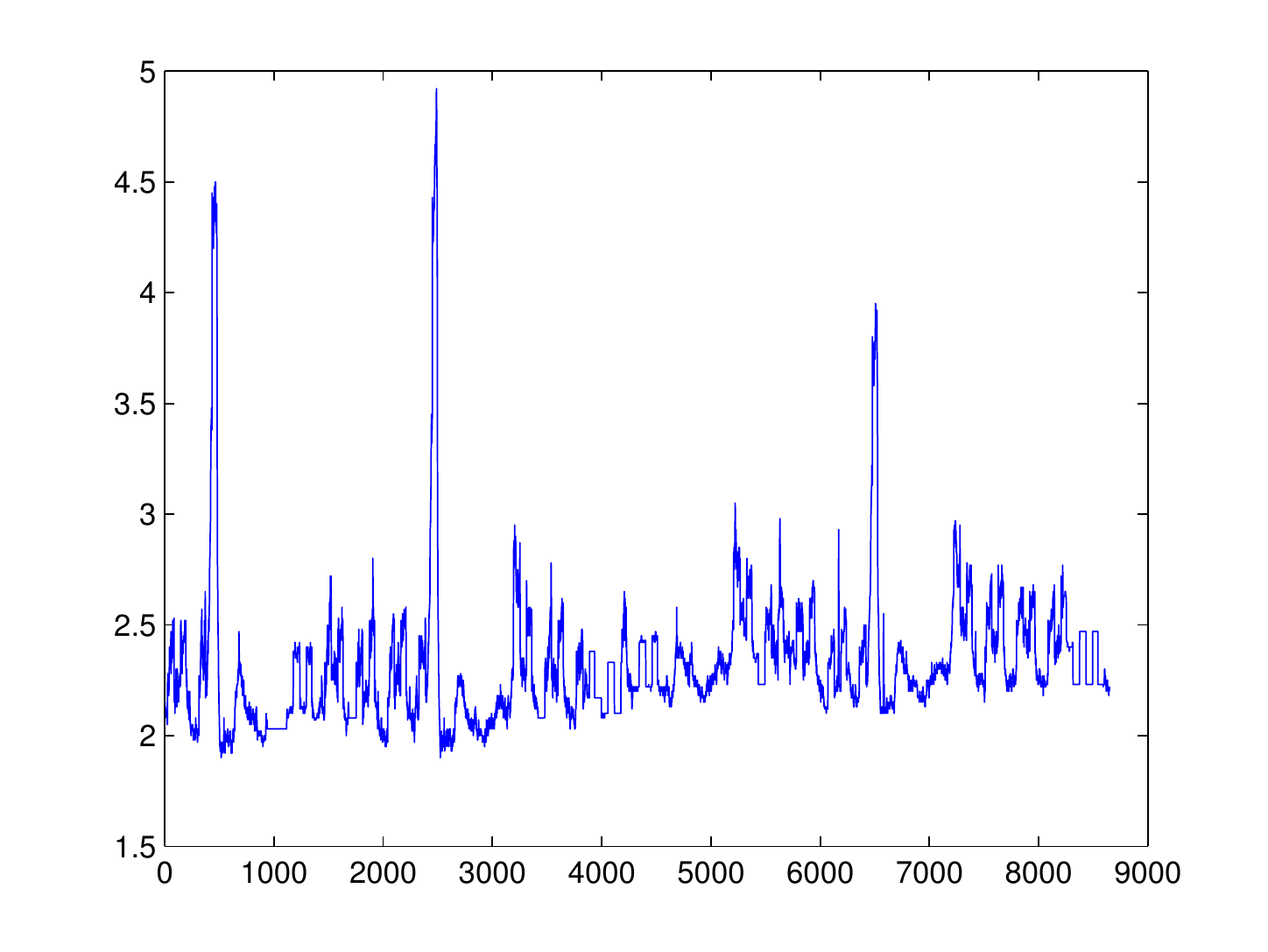}}\quad
\bigskip
\bigskip
\caption{Experiment Network and Sample Travel Time Data}
\label{fig:5-network}
\end{center}
\end{figure}
\begin{enumerate}
 \item
A regression estimate is generated for the trend over the year in
the table, and the estimate is not significant except as a stable
mean function $a=2.23$. This part is shown in the Figure
\ref{fig:80edetrend}.

\item

The seasonal model is studied using a regression method. The
parameters for the two seasonal models are obtained as follows with
T as 2064 for the weekly pattern and 288 for the daily pattern. The fitted parameters are displayed in Table \ref{fig:80edeseasonal} and the
seasonal parts are shown in Figure \ref{fig:80edetrend}.
\begin{table}[ht]
\caption{Seasonal effects in the 80E data} \label{fig:80edeseasonal}
\centering
\begin{tabular}{|c|c|c|c|c|c|c|c|c|}
\hline
Weekly trend& $k$ &$a_1$ &$a_2$ &$a_3$ &$b_1$& $b_2$&$b_3$&T\\
  &0.0927  & -0.0135 &   0.0794 &  -0.0902 &   0.0477 &  -0.0873 &  -0.0228& 2064\\

% &-0.293 &1.241 & 1.362 & 0.387 & -0.268 & 0.377 & 0.631& 2064\\
 \hline
Daily trend&&&&&&\\
%&-0.206 &-0.664 & 0.728 & 0.399 & -2.236 & 0.0749& -0.141&288\\
& 0.0002&    0.0240&    0.1143&   -0.0795&   -0.1509& 0.0127&-0.0204&288\\
\hline
\end{tabular}
\end{table}

\begin{figure}[ht]
\begin{center}
\subfloat[Mean function over time]% \quad on the next line adds spacing
{\includegraphics[angle =0 ,width=0.45
\textwidth]{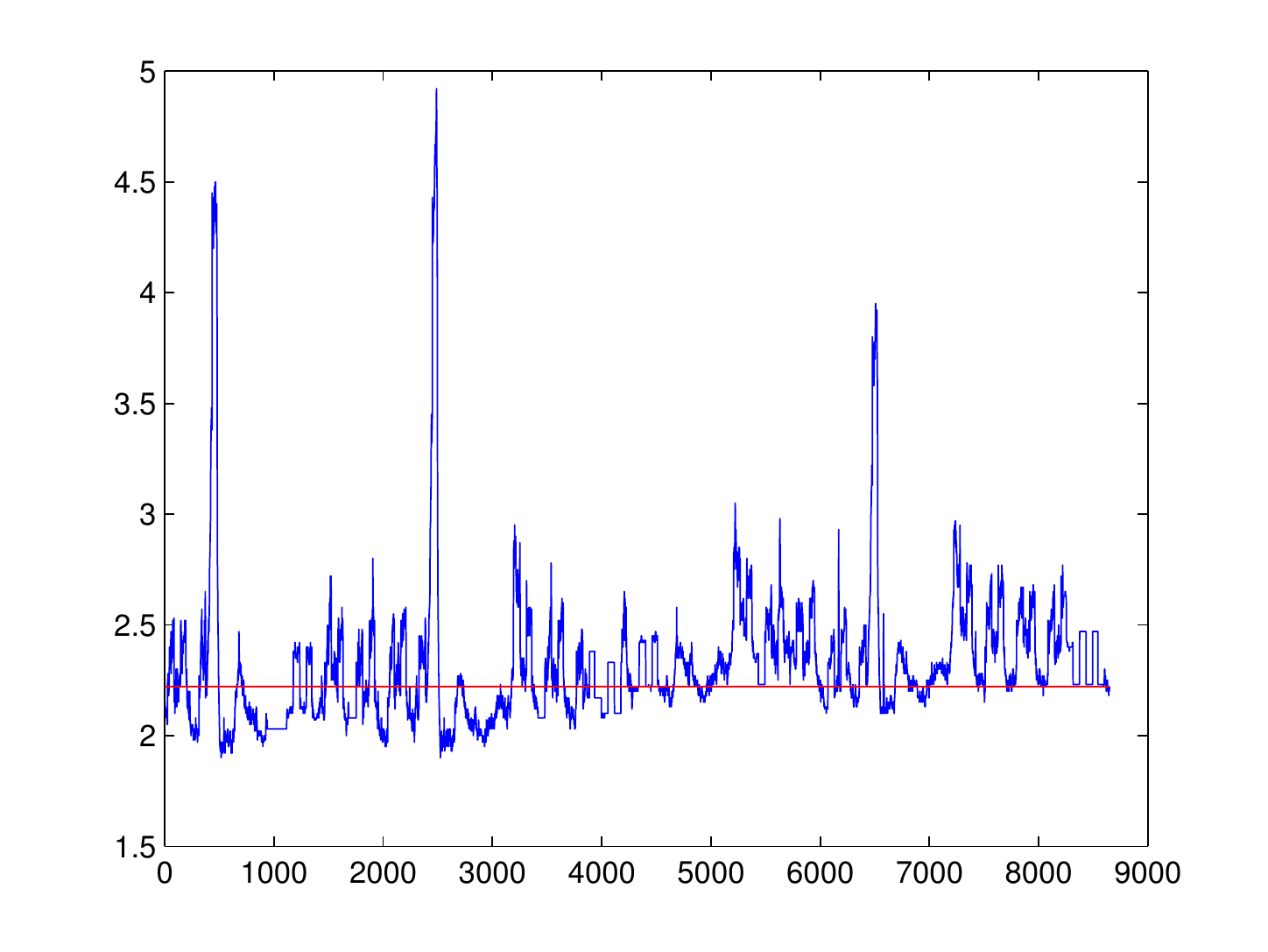}}\quad
\bigskip
\bigskip
\subfloat[The weekly pattern] {\includegraphics[angle=0 , width=0.45
\textwidth]{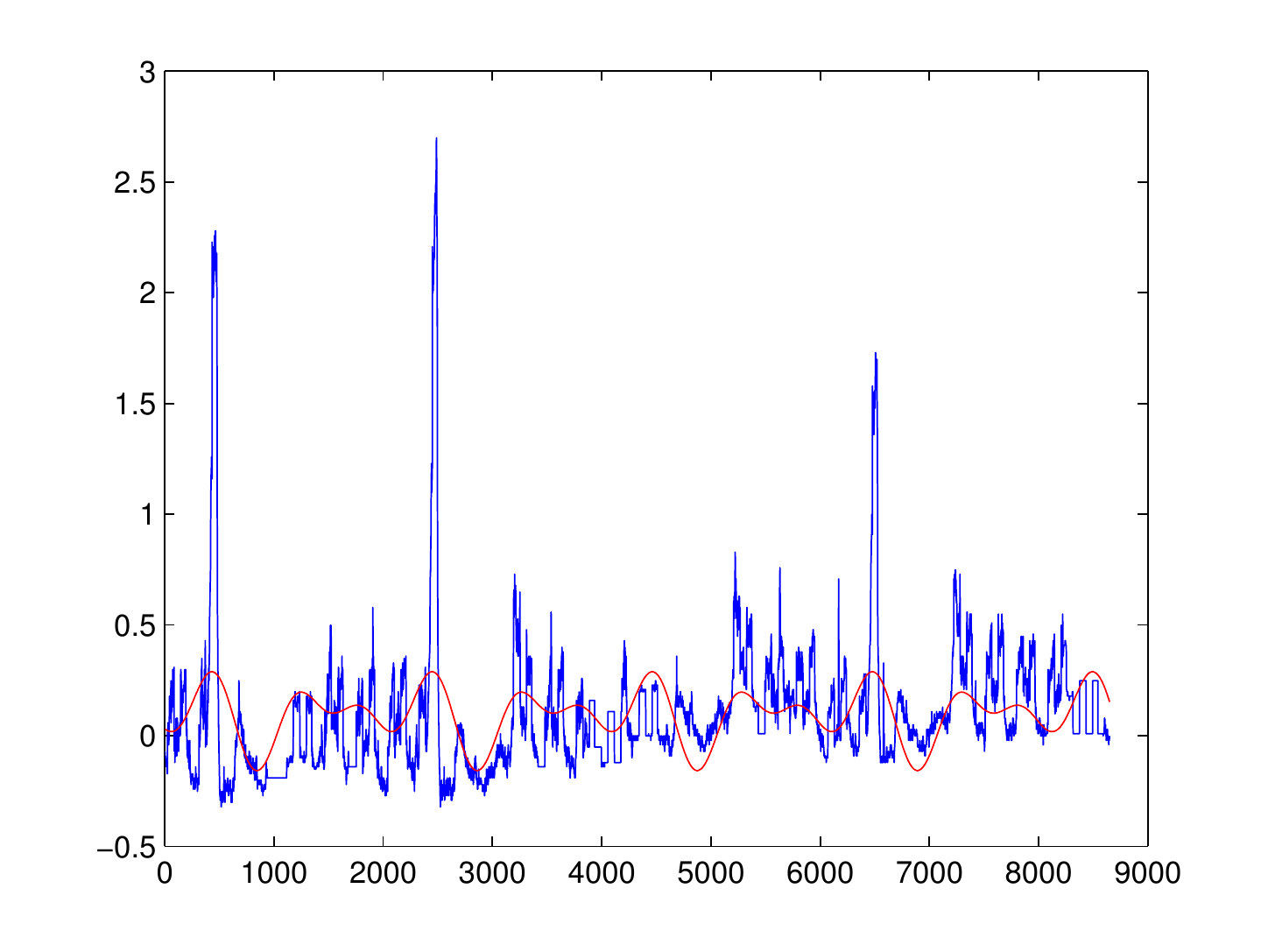}}\quad
\bigskip
\bigskip
\subfloat[The daily pattern]% \quad on the next line adds spacing
{\includegraphics[angle =0 , width=0.45
\textwidth]{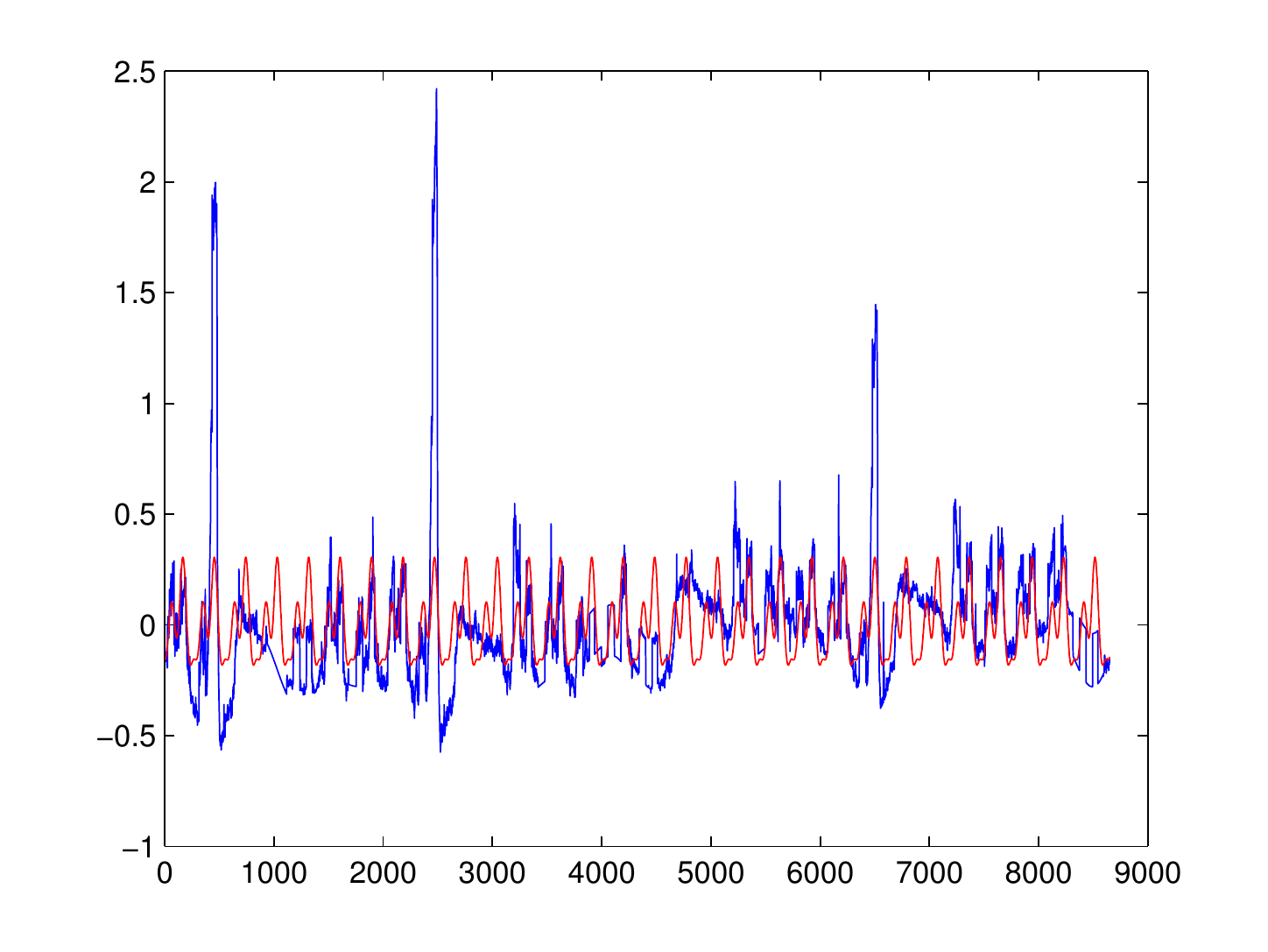}}\quad
\bigskip
\bigskip
\subfloat[The residual] {\includegraphics[angle=0 , width=0.45
\textwidth]{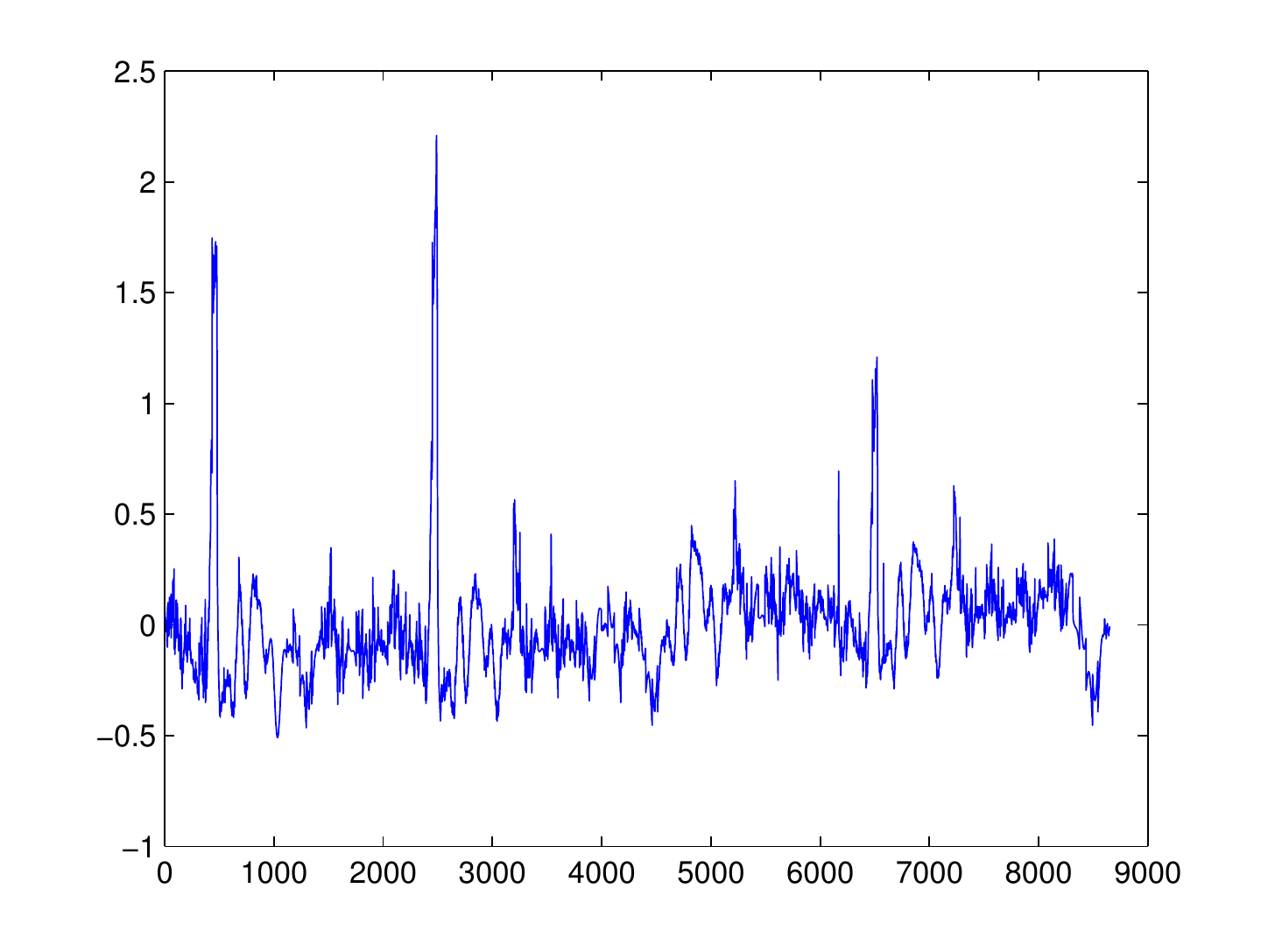}}\quad
\bigskip
\bigskip
\caption{Remove the trend, weekly, daily part of 80E data series}
\label{fig:80edetrend}
\end{center}
\end{figure}

\item After removing the trend and seasonal part, the residual is used to estimate the CARMA($p,q$) process.
In the literature, the methods to estimate CARMA($p,q$) models can
be based either directly on the continuous-time process or on a
discretised version. The latter relates the continuous-time dynamics
to a discrete time ARMA process. The advantage of this method is
that standard packages for the estimation of ARMA processes may be
used in order to estimate the parameters of the corresponding CARMA
process. However, not every ARMA($p,q$) process is embeddable in a
CARMA($p,q$) process. Brockwell and collaborators devote several
papers to the embedding of ARMA processes in a CARMA process,
\cite{brockwell1995note} and \cite{brockwell1999class}. In the study
below, this approach is employed by assuming an appropriate class of
CARMA processes to work with.

Following the intuition above, the residual is identified as the
following ARIMA(1,1,0) model (Table \ref{tab:5-dsde}). A Box-Ljung
test suggests that the p-value $=0.1928$. This model describes the
time series well.

%$$(T_{t}-T_{t-1})=a(T_{t-1}-T_{t-2})+b \sigma e_t+\sigma e_{t-1}$$

%\begin{table}[htm]
%\caption{ARIMA modeling of the residuals} \label{tab:5-dsde}
%\centering
%\begin{tabular}{|c|c|c|c|}
%\hline
%Weekly trend& a & b & $\sigma$\\
%value & 0.1197 & 0.2612&0.265\\
% s.e. & 0.0281& 0.0272&0.265\\
%\hline
%\end{tabular}
%\end{table}

\begin{figure}[ht]
\begin{center}
\includegraphics[angle=0, width=0.7 \textwidth]{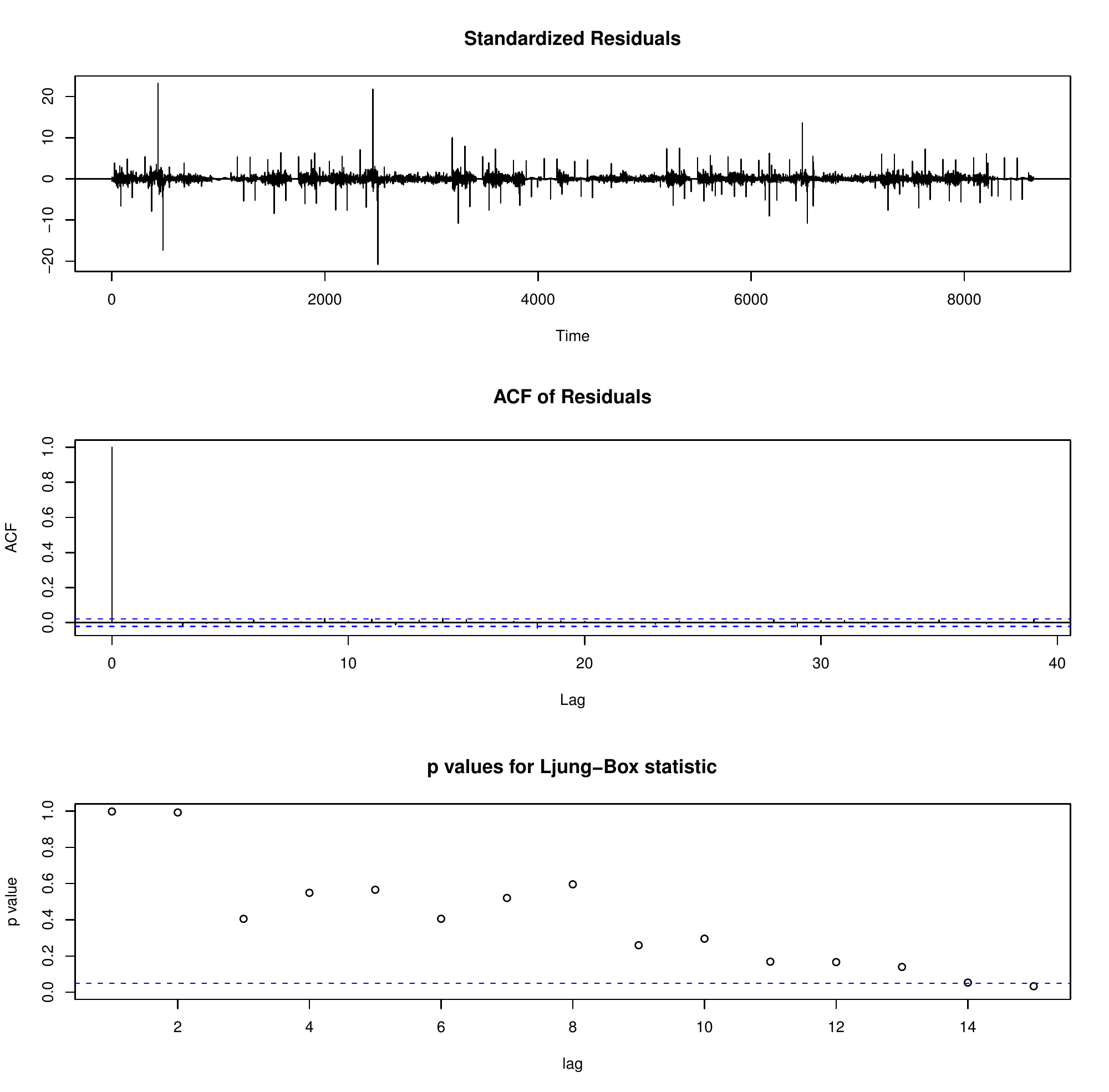}
\caption{80E ARIMA model} \label{fig:80earima}
\end{center}
\end{figure}

\begin{table}[ht]
\caption{Model selection for 80E data} \label{tab:5-dsde} \centering
\begin{tabular}{|c|c|c|c|}
\hline
&Log Liklihood&AIC&p-value of Box-Ljung test\\
 ARIMA($1,0,0$)&14467.36&-28928.72& 0.0035\\
 ARIMA($1,1,0$)&14438.3& -28872.61&0.1928\\
 ARIMA($1,1,1$)&14438.31& -28870.61&0.1322\\
 ARIMA($2,1,1$)&14438.31&-28868.61&0.0840\\
 %ARIMA($1,1,1$)-Garch($1,1$)&15784.7&&0.3043\\
% ARFIMA($1,0.49,1$)&&&9&$<$ 2.2e-16\\
\hline
\end{tabular}
\end{table}

 The model is described as the ARIMA($1,1,0$) model by comparing the test
 results, and additional variance modeling can be conducted to describe the volatility
 process.

\item
The estimation is demonstrated for a selected path above, in which
the model follows ARIMA($1,1,0$). However, travel time processes for
different paths may appear to fit ARIMA models with different orders
(Table \ref{tab:5-bfdl}). To address such variations, the pricing of
travel time derivatives is discussed considering all possible orders
of ARIMA models. Particularly, distinct treatments are applied
conditioned on whether $p > q$ holds or not. In more detail, the
spectral density for a CARMA($p,q$) process is defined as

$$f(\omega)=\frac{\sigma^2\beta(i\omega)\beta(-i\omega)}{2\pi\alpha(i\omega)\alpha(-i\omega)}$$

The stationarity condition of the CARMA processes requires the roots
of the polynomial $\alpha(z)$ to have negative real-parts and that
$p > q$, \cite{tomasson2011some}. Below the estimation of
CARMA($p,q$) processes are conducted based on fitted ARMA model
considering different relations of $p$ and $q$:

\begin{table}[ht]
\caption{Models with best fit for different paths}
\label{tab:5-bfdl} \centering
\begin{tabular}{|c|c|c|}
\hline
Path Name&Model&p-value of Box-Ljung test\\
Cummings Skwy - Maritime Academy&ARIMA($1,1,0$)& 0.1928\\
Berkeley-Davis&ARIMA($2,1,1$)& 0.0503\\
Lincoln-Davis&ARIMA($1,1,1$)& 0.3419\\
\hline
\end{tabular}
\end{table}

\begin{enumerate}

\item If $p>q$, the discrete solution for the CARMA($p,q$) state space
representation can be given as the following equation
$$x_{t+\delta t}=e^{A\delta t}x_t+\int_t^{t+\delta t}e^{t+\delta t-u}e_p\sigma_udB_u$$
The noise term is of variance $\int_0^{\delta t}
e^{Au}e_pe_p^te^{A^tu} du$. If first order expansion of the matrix
exponential $e^{Au}$ is taken, a set of formulas can be derived to
map the parameters of the CARMA($p,q$) processes
($\alpha_n$,$\beta_n$) approximately to the parameters of the
corresponding discrete ARMA($p,q$) models ($a_n$,$b_n$),
\cite{hardle2012implied}. Correspondingly, the parameters of the
CARMA($p,q$) processes can be backed out using estimated ARMA model
parameters:
%The formulas can be obtained
%by substituting the solution to the CAMRA($p,q$) processes into the
%discrete-time dynamics and
 Assuming the grid size is $h$, the mapping for the auto regression parts of the model in the first three
orders is displayed in Table \ref{tab:5-darmacarma}.

\bigskip
\bigskip
\bigskip
\bigskip

\begin{table}[ht]
\caption{Coefficients mapping between discrete and continuous ARMA
models} \label{tab:5-darmacarma} \centering
\begin{tabular}{|c|c|}
\hline
Model&Formula\\
CAR(1) &$a_1=1-\alpha_1h$\\
\hline
CAR(2) &$a_1=2-\alpha_1h$\\
 &$a_2=\alpha_1h-\alpha_2h-1$\\
%\hline
%CAR(3) &$a_1=3-\alpha_1h$\\
% &$a_2=2\alpha_1h-\alpha_2h-3$\\
%&$a_3=-\alpha_1h+\alpha_2h-\alpha_3h+1$\\
%\hline
%CARMA($p,q$) ($p>q$)&$b_n=\beta_n$\\
\hline
\end{tabular}
\end{table}

The estimation of the moving average parameters can be based on auto
correlation function \cite{espen2012futures} and a least absolute
deviation algorithm can be used to estimate the MA part based on the
empirical and theoretical autocorrelation functions of the CARMA
processes.
$$\gamma(s)=b^te^{A|s|}\Sigma b$$
where
$$\Sigma=\int_0^\infty e^{Au}e_pe_p^te^{A^tu}du=-A^{-1}e_pe_p^t$$
In this representation, $A^{-1}$ is the inverse of the operator A :
$X \rightarrow AX + XA^t$, \cite{pigorsch2009multivariate} and
\cite{barndorff2007positive}.

% by considering the linearity
%of the model. Taking the CARMA($2,1$) for example, the calculation
%is conducted based on the following mapping formula which holds for
%CARMA($2,0$) process, \cite{hardle2012implied} ��

%$$X_{1(t+2)} \approx (2 -\alpha_1 )X_{1(t+1)} + (\alpha_1 -
%\alpha_2 + 1)X_{1(t)} + \sigma_t(B_{t-1} -B_t)$$
%
%Then taking an additional time lag on both sides of this equation:
%$$X_{1(t+1)}\approx(2 -\alpha_1 )X_{1(t)} + (\alpha_1 -
%\alpha_2 + 1)X_{1(t-1)} + \sigma_t(B_{t-2} -B_{t-1})$$

%For a CARMA($2,1$) which shares the same AR coefficients as the
%CARMA($2,0$) processes above, the two formulas above can be
%aggregated using the moving average coefficients $b_0$ and $b_1$:
%
%\begin{eqnarray*}
%b_0X_{1(t+2)}+b_1 X_{1(t+1)}&\approx&(2 -\alpha_1
%)(b_0X_{1(t+1)}+b_1
%X_{1(t)})+ (\alpha_1 - \alpha_2+1)\\
%&& (b_0X_{1(t)}+b_1X_{1(t-1)})  + b_0(\sigma_t(B_{t-1} -B_t))\\
%&&+b_1(\sigma_t(B_{t-2} -B_{t-1}))\\
%X'_{1(t+2)}&\approx&(2 -\alpha_1 )X'_{1(t+1)} + (\alpha_1 -
%\alpha_2 + 1)X'_{1(t)}\\
%&& + b_0\sigma_t(B_{t-1} -B_t)+b_1\sigma_t(B_{t-2} -B_{t-1})\\
%\end{eqnarray*}
%
%Then this formula suggests that CARMA($2,1$) processes can be
%approximated using the discrete representation. This mapping holds
%when $b_0=1$; if $b_0\neq 1$ for the continuous CARMA($p,q$)
%process, a normalization can be conducted first.

With these mapping formulas, the coefficients of the continuous ARMA
processes can be solved using the parameters of the fitted discrete
ARMA model. These mappings ignore higher order terms and can hence
be applied to low order CARMA processes for engineering purposes.
These values can also be used as the initial values for more precise
estimation of the CARMA parameters.
% This is sufficient for the application here as

 To obtain such precise estimations, a Kalman filter is used to further extract the unobserved states of the
CARMA ($p,q$) processes based on the state-space representation and
parameters are estimated by minimizing the error between observed
and estimated output series of the state space model. In more
detail, given a travel time series whose trend and seasonality are
removed, first order differentiation is taken to remove the
integrated part of the model. The remaining process $X_t$ is then
processed using a Kalman filter to extract the unobserved higher
order states in the CARMA process, \cite{brockwell2011estimation}
and \cite{Eckhard2012estimation}. The filtered model produces an
output series $X'_t$ and the mean square error of $X_t$ and $X'_t$
are minimized to yield the minimum mean-squared-error linear
predictors for the parameters of CARMA($p,q$), which is found by
numerically minimizing the sum of squares. To obtain a warm start,
the parameters obtained from the previous first order approximation
are used as initial values. The estimated parameters are displayed
in Table \ref{tab:5-carmaest} and the estimated processes need to be
calibrated further to the market to determine the market value of
risk $\theta$, using the methods discussed in next section.

\begin{table}[ht]
\caption{Coefficient estimation for discrete and continuous time
ARMA models} \label{tab:5-carmaest} \centering
\begin{tabular}{|c|c|c|c|c|}
\hline
Path&Parameter&ARMA&Initial Value&Estimated Value\\
\hline
Cummings Skwy - &AR-1&-0.0326&0.2065 &0.2512\\
Maritime Academy&&&&\\
\hline
Berkeley-Davis&AR-1&0.9646&0.2071 &0.1780\\
&AR-2&-0.1275&0.03258 &0.0685\\
&MA-1&-0.6575&0.1356 &-0.6296\\
%CAR(2) &$a_1=2-\alpha_1h$\\
% &$a_2=\alpha_1h-\alpha_2h-1$\\
%\hline
%CAR(3) &$a_1=3-\alpha_1h$\\
% &$a_2=2\alpha_1h-\alpha_2h-3$\\
%&$a_3=-\alpha_1h+\alpha_2h-\alpha_3h+1$\\
%%\hline
%%CARMA($p,q$) ($p>q$)&$b_n=\beta_n$\\
\hline
\end{tabular}
\end{table}

\item
For the processes corresponding to the case of $p \leq q$, for
example, the ARIMA($1,1,1$) on path Lincoln-Davis, it is recommended
that Monte Carlo method based on the discrete ARIMA model be used to
price the travel time derivative after the pricing principles and
risk neutral measure are specified. Although CARMA ($p,q$) has
recently been extended into generalized random processes for cases
when $p \leq q$, \cite{brockwell2010carma}, the processes under such
settings are not well defined in the usual sense. In terms of
rationale, the general random process yields a way to represent the
discrete approximation of the high order derivatives of Ornstein -
Uhlenbeck process with respect to selected linear functionals, but
this representation does not lead to typical continuous time
stochastic models. The application of a generalized random process
for pricing travel time derivatives is still contingent upon further
studies of such processes and hence the Monte Carlo method is a
reasonable approximation for modeling and pricing travel time
derivative.

\end{enumerate}
\end{enumerate}

In summary, travel time processes are fitted to discrete ARMA($p,q$)
models with trend and seasonality adjustment first. Then the fitted
models are converted into stable CARMA($p,q$) processes for further
calculation when $p > q$. When $p \leq q$, it is recommended that
Monte Carlo simulation based on the discrete ARMA models be used to
price travel time derivatives.
%while the representation of
%generalized random process may lead to further theoretical asset
%pricing in a generalized sense.
In the following sections, potential pricing schemes are discussed
based on fitted CARMA($p,q$) processes.

\subsection{Risk neutral pricing in an incomplete market}

To obtain a general pricing expression for travel time derivatives,
risk neutral pricing principle in incomplete market conditions is
employed in this section, deploying the method in
\cite{benth2007volatility}.

\subsubsection{Risk neutral representations}

%The risk-neutral probability $Q$.
 A risk-neutral probability is by definition
a probability measure $Q \sim P$ such that all tradable assets in
the market are martingales after discounting. Thus, all equivalent
probabilities $Q$ will become risk-neutral probabilities. A
sub-family of probability measures $Q$ is specified using the
Girsanov transformation: assume $\omega_t$ is a real-valued
measurable and $\frac{\omega_s}{\sigma_s}$ is a bounded function.
The stochastic process:

$$Z^\omega(t)=exp(\int \frac{\omega_s}{\sigma_s}dW_s-\frac{1}{2}\int \frac{\omega^2_s}{\sigma^2_s}ds)$$
is the density process of the probability measure $Q$. Under $Q$,
the process
$$dB_t=dW_t-\frac{\omega(s)}{\sigma(s)}dt=dW_t-\theta_t$$
is a Brownian motion and $\theta_t$ is called the market value of
risk process. Based on this measure change, the dynamics of the
underlying process under the risk neutral measure are given by the
following S.D.E:

$$dT_t=d c_t+ \omega(t)dt+a_t(c_t-T_t)dt+\sigma_t dW_t$$

and the solution to this S.D.E is then

$$T_t= c_t+(T_0-c_0)e^{-\int_0^t a_s ds}+e^{-\int_0^t a_s ds}\int_0^t e^{\int_0^u a_s ds} \omega_u du + e^{-\int_0^t a_s ds} \int_0^t e^{\int_0^u a_s ds}\sigma_u dW_u$$

When the coefficients are constant, the general solution reduces to
the following simpler form:

$$T_t= c+ (T_0-c)e^{-at}+ \int_0^t e^{-a(t-u)} \omega_u du+ \int_0^t e^{-a(t-u)}\sigma dW_u$$

Then the prices of different travel time derivative contracts in
continuous time are calculated as the discounted expectation of
their corresponding payoff function under this measure, which is
given below:

The price of a HCD futures contract can be calculated as:
$$ F_{HCD}(t_1,t_2,T,t)=E_Q\left\{\int_{t1}^{t2}max(T_s-T,0) ds
|\mathcal{F}_t\right\}$$

The price of a LCD futures contract can be calculated as:
$$
F_{LCD}(t_1,t_2,T,t)=E_Q\left\{\int_{t1}^{t2}max(T-T_s,0) ds
|\mathcal{F}_t\right\}$$

Assuming a constant interest rate, the price of a HCD call options
contract can be calculated as:
%$$C_{HCD}(t_1,t_2,T,t)=e^{-rt}E_Q\left\{max(F_{HCD}(t_1,t_2,T,t)-K,0)|F_t\right\}$$
$$C_{HCD}(t_1,t_2,T,t,K)=e^{-rt}E_Q\left\{max(\int_{t1}^{t2}max(T_s-T,0) ds-K,0)|\mathcal{F}_t\right\}$$

and  the price of a LCD call options contract can be calculated as
%$$C_{LCD}(t_1,t_2,T,t)=e^{-rt}E_Q\left\{max(F_{LCD}(t_1,t_2,T,t)-K,0)|F_t\right\}$$
$$C_{LCD}(t_1,t_2,T,t,K)=e^{-rt}E_Q\left\{max(\int_{t1}^{t2}max(T-T_s,0) ds-K,0)|\mathcal{F}_t\right\}$$

The price of a CTT futures contract can be calculated as:
$$ F_{CTT}(t_1,t_2,t)=E_Q\left\{\int_{t1}^{t2}T_s ds |\mathcal{F}_t\right\}$$

Assuming a constant interest rate, the price of a CTT call options
contract can be calculated as
%$$C_{CTT}(t_1,t_2,t)=e^{-rt}E_Q\left\{max(F_{CTT}(t_1,t_2,t)-K,0)|F_t\right\}$$
$$C_{CTT}(t_1,t_2,t,K)=e^{-rt}E_Q\left\{max(\int_{t1}^{t2}T_s ds
-K,0)|\mathcal{F}_t\right\}$$

%The dynamics of these options can be complicated. The option price
%can be found by simulating the travel time process under the
%specific measure $Q$.
Since travel time is neither a tradable nor storable asset, the
derivatives contracts cannot be hedged using travel time itself in
the financial markets, and the market of the travel time derivatives
is therefore incomplete. Under such incomplete markets, the risk
neutral measure is not unique. To obtain the prices, the risk
neutral measure should first be specified considering the
characteristics of the incomplete market setting. Moreover, the
expectation can be calculated using two alternative methods: it can
be calculated using Monte Carlo simulation of the underlying
processes under the specific risk neutral measure, the average
discounted payoff of the financial derivative in all paths yield the
price of the contract; alternatively, some explicit formulas can be
obtained by considering the property of the discounted price
processes under the risk neutral measure.

\subsubsection{Determination of the risk neutral measure}

The incompleteness of the travel time derivative market requires the
estimation of the market price of risk (MPR) for pricing and hedging
travel time derivatives. The market price of risk adjusts the
underlying process representing travel time so that the implied
price is arbitrage free.
% If the process is driven
%by a Brownian motion, then ordinary martingale measure can be used
%to price the derivatives by specifying the market value of risk
%process
% $\Omega_t=exp(-\frac{1}{2}\theta_t^2-\theta_t B_t )$
 As it is stated in Section 5.4.3, $B_t=W_t-\theta_t$ is a Brownian motion under the risk neutral measure.
 Since the underlying asset is not tradable, there is no unique risk
neutral measure.
%In general, derivative pricing in an incomplete
%market relies on the positions of the underlying asset and the
%position of the option itself \cite{DennisIncom}.
The drift of the asset price process is the view of the trader about
the growth of the process.
%Since the travel time process is almost independent from the
%interest rate process, this assumption can be well justified.
 There are different ways of specifying this measure and some of them are discussed
below:

% As it is noted, the market is incomplete due to that travel time is not a
%tradable assets.

\begin{enumerate}
\item Market value of risk can be inferred from market traded products.
\cite{hardle2012implied} suggests inferring the market price of
risk(MPR) from traded CAT futures by minimizing the mean square
error between the modeled contract prices with the market traded
prices. Once the MPR for temperature futures is known, it is used to
price other derivatives. According to \cite{meyer2008multi}, within
incomplete markets, there may exists many equivalent risk-neutral
measures; it is the job of the market as a whole, via trading of
derivatives, to decide which measure prevails at any one given point
in time. A class of equivalent martingale measures can be identified
which maintains the structure of real-world dynamics for asset
prices. These measures can then be used to obtain forward prices and
value spread option. Moreover, the differences in pricing measures
leads to risk premium which can be calibrated using market prices:
\cite{lucia2002electricity} showed a negative market price of risk
associated to the non-stationary term in their two-factor models,
when analyzing data from energy market. In the proposed two-factor
model, where the non-stationary term is a drifted Brownian motion,
the negative market price of risk appears as a negative risk-neutral
drift. \cite{benth2008pricing}, showed that using the certainty
equivalence principle that the presence of jumps in the spot price
dynamics will lead to a positive risk premium in the short end of
the futures curve. \cite{bessembinder2002equilibrium} explain the
existence of a positive premium in the short end of the futures
market by an equilibrium model.

In the context of travel time derivatives, the choice of $\theta_t$
uniquely determines the equivalent martingale measure under which
derivatives pricing is performed. One way of defining the market
price of risk is to extrapolate from option prices. This technique
resembles recovery of the implied volatility in the Black-Scholes
model. A chosen objective function can be minimized to find
$\theta_t$, such as the mean absolute percentage error between the
market and model option prices. The market prices can be chosen as
averages of the bid and ask offers and options with different
strikes can be used to calibrate $\theta_t$ for a given day.
Alternatively, calibrating it to futures prices is also feasible.
The procedure  is analogous to that used with options, and the model
can be calibrated to one futures price. Futures have more liquidity
than options and hence allow for a more frequent and precise
calibration of $\theta_t$. The value of $\theta_t$ is subject to the
incentive for hedging on the demand side relative to the supply
side. For a concrete example, the market value of risk process
$\theta_t$ can be calibrated by a set of HCD futures contracts based
on the common underlying travel time series by minimizing the mean
squared difference between modeled price and traded price below

$$\theta_t=argmin(\sum_i(F_{i,market}-F_{i,HCD}(t_1,t_2,T,t,\theta_t))^2)$$

Different contracts can be used to conduct such calibration and
contracts with the most liquidity in the market are the best
instrument for such purposes.

\item Suitable hedging strategies result in a price, that suggests a risk neutral measure.
  Different hedging strategies leads to varying derivative prices.
\item Some characteristics of the risk neutral measure can be specified based on
  certain optimality conditions. For example, the
minimum entropy measure has
   been studied in
 \cite{frittelli2000minimal}, as it can yield reasonable asset
 prices and there is a huge literature on the use of
maximum entropy measure for calibration purposes.
%Other martingale measure can also be identified
%(\cite{artzner1995approximate}).
In \cite{follmer1991hedging}, a minimal martingale measure based on
local variance minimization provides a strategy that penalizes
over-hedging. \cite{xu2006risk} introduces pricing methods based on
suitable risk measures and partial hedging is used when certain risk
measures are introduced to control the residual risk at expiration.
Such methods can be applied to identify the best pricing measure for
pricing travel time derivatives.
\end{enumerate}

%This paper introduces the financial derivative based on travel
%time, which is based on road travel times in transportation systems.
%It is a new value pricing scheme based not only on the level of
%congestion but also on the uncertainties surrounding the congestion.
%The product design and pricing of travel time derivatives is a
%meaningful and promising joint research topic shared by
%transportation research and financial engineering fields.
%
These methods may lead to different prices due to the non-tradable
nature of travel time. This discussion again shows that the prices
of travel time derivatives are subject to the choices of specific
hedging strategies, under incomplete market conditions with
non-unique risk neutral measures. In the following section, it is
assumed that a risk neutral measure can be calibrated using the
price of traded derivatives.

\subsubsection{Risk neutral pricing for travel time derivatives}

In this section, pricing P.D.Es are further derived based on the
pricing measure that is identified using methods in the previous
section. The rationale is that derivative prices are functions of
underlying processes and these prices should be martingales under
the specified risk neutral measure. To provide some background, a
martingale is a stochastic process for which, at a particular time
in the realized sequence, the expectation of the next value in the
sequence is equal to the present observed value even given knowledge
of all prior observed values at the current time. Based on this
rationale, Proposition 8.1 of \cite{steele2001stochastic} introduces
the martingale P.D.E condition for a stochastic process, which
suggests the drift term should be zero for a martingale process. As
the mathematic derivatives of the price function can be calculated,
the martingale condition above leads to partial differential
equations (P.D.E) which yields the analytical solution for the
prices of travel time derivatives. In the following analysis, the
prices based on simple CARMA($1,0$) processes with first order
integration are first discussed and then extended to general
CARMA($p,q$) processes with first order integration and with $p
> q$; $Y_t$ is obtained by removing trends and seasonal adjustments
from the original travel time series.

%\begin{thm}
% For European call options based on the process $Y_t$, where $Y_t=\int_0^t
%X_udu$ and $X_t$ is a CARMA($1,0$) process, the price can be found
%via the following P.D.E, after specifying a risk neutral measure
%$P$:
%
%$$v_t(t,x,y)+A(t,X_t)v_x(t,x,y)+xv_y(t,x,y)+\frac{1}{2}B(t,X_t)^2v_{xx}(t,x,y)=rv(t,x,y)$$
%$0 \leq t < T$,$x \in R$,$y \in R$
%%\newnot{symbol:v}
%
%with the following boundary conditions:
%
%$$v(t,0,y)=e^{-r(T-t)}(y-K)^+, 0 \leq t < T, y \in R$$
%$$lim_{y\rightarrow -\infty} v(t,x,y)=0,0 \leq t < T, x \in R$$
%$$ v(T,x,y)=(y-K)^+, x \in R, y \in R$$
%\end{thm}
%
%proof: Due to the existence of integration,i.e. $Y_t=\int_0^t
%X_udu$, to price a European call option on $Y_t$ is similar to
%pricing an Asian call option on the primary process $X_t$. The proof
%follows from Theorem 7.5.1 of \cite{shreve2004stochastic}. As $Y_t$
%is obtained by removing trends and seasonal adjustments from the
%original travel time series, it can be negative and the boundary
%conditions are slightly different.

\begin{thm}
For the Asian type call option on the process $Y_t$, where
$Y_t=\int_0^t X_udu$ and $X_t$ is a CARMA($1,0$) process, its price
can be solved via the following P.D.E by further expanding states as
follows:
$$v_t(t,x,y,z)+A(t,X_t)v_x(t,x,y,z)+xv_y(t,x,y,z)+yv_z(t,x,y,z)+\frac{1}{2} B(t,X_t)^2 v_{xx}=rv(t,x,y,z)$$
$0 \leq t < T$,$ x  \in R $,$y \in R$,$z \in R$

with boundary conditions:

$$lim_{z\rightarrow -\infty} v(t,x,y,z)=0, 0\leq t<T,x \in R, y \in R$$
$$ v(T,x,y,z)=(z-K)^+, x \in R, y \in R, z \in R$$
$$v(t,0,0,z)=e^{-r(T-t)}(z-K)^+, t<T, z\in R$$

\end{thm}

Proof:

 Consider the claim $(Z_t-K)^+$, under the
risk neutral measure $Q$, we have:
\begin{eqnarray*}
dZ_t&=&Y_tdt\\
 dY_t&=&X_tdt\\
 dX_t&=&A^*(t,X_t)dt+B(t,X_t)dW_t\\
\end{eqnarray*}
We consider that this group of differential equation defines a three
dimensional Markovian process. The value of the derivative contract
will be $P=de^{-rt}v(t,x,y,z)$. By Ito'lemma, it is subject to the
following dynamics:
\begin{eqnarray*}
dv(t,x,y,z)&=&v_t dt+v_x dx+\frac{1}{2}v_{xx}d<x>+v_ydy+v_zdz\\
dP&=&e^{-rt}(-rv+v_t+\frac{1}{2}v_{xx}B(t,X_t)^2+ v_x
A^*(t,X_t)+xv_y+yv_z)dt+ v_x
B(t,X_t)e^{-rt}dW_t\\
\end{eqnarray*}
The discounted price process should be a martingale under the risk
neutral measure, so we have
\begin{eqnarray*}
-rv+v_t+\frac{1}{2}v_{xx}B(t,X_t)^2+ v_x A^*(t,X_t)+xv_y+yv_z=0\\
\end{eqnarray*}
 That is
\begin{eqnarray*}
v_t+\frac{1}{2}v_{xx}B(t,X_t)^2+ v_x A^*(t,X_t)+xv_y+yv_z=rv\\
\end{eqnarray*}
Here $A^*(t,X_t)$ is the drift of $X$ under the risk neutral
measure.

% As $X$ is not tradable, it is not necessarily $r$, the primary
%drift $A(t,X_t)$ can be used for certain pricing measure $Q$

For the boundary conditions, as $Y_t$ is obtained by removing trends
and seasonal adjustments from the original travel time series, it
can be negative. All the state variables corresponding to the
CARMA($1,0$) process can be in $R$.

If $Y(t)$ approaches $-\infty$, then the probability that the call
expires in the money approaches zero and the option price approaches
zero. This leads to the first boundary condition. The second
boundary condition is just the payoff of the call option.

The pricing of european call options can be derived in a reduced
form.
 Q.E.D.

 Noticing the similarity in defining the state space representation of the CARMA($p,q$) and the Asian option pricing formula above,
 additional state expansion is employed to price the travel time derivatives based on CARMA($p,q$) processes.
 For example, consider the Asian option based on $Y_t$, where $Y_t=\int_0^t
X_udu$ and $X_t$ is a CARMA($2,0$) process %the CARMA($2,1,0$)
 %processes
 , the state space dynamic can be described by the
 following set of S.D.E:
\begin{eqnarray*}
dZ_t&=&Y_tdt\\
 dY_t&=&X^0_tdt\\
 dX^0_t&=&X^1_tdt\\
 dX^1_t&=&A^*(t,X^0_t,X^1_t)dt+B(t,X^0_t,X^1_t)dW_t\\
\end{eqnarray*}

where $X^0_t$ and $X^1_t$ construct the CAR($2$) process, and $Y_t$
represents the CARMA($2,0$) process, and $Z_t$ characterizes the
integrated price process in the payoff function of the Asian option.
This group of differential equations defines a four dimensional
Markovian process. The value of derivative contract will be
$de^{-rt}v(t,x_1,x_0,y,z)$. By Ito'lemma, it is subject to the
following dynamics:
\begin{eqnarray*}
dv(t,x_1,x_0,y,z)&=&v_t dt+v_{x_1} d{x_1}+\frac{1}{2}v_{x_1x_1}d<x_1>+v_{x_0}dx_0+v_ydy+v_zdz\\
dP&=&e^{-rt}(-rv+v_t+\frac{1}{2}v_{x_1x_1}B(t,X^0_t,X^1_t)^2+
v_{x_1} A^*(t,X^0_t,X^1_t)\\&&+x_1v_{x_0}+x_0v_y+yv_z)dt+ v_{x_1}
B(t,X^0_t,X^1_t)e^{-rt}dW_t\\
\end{eqnarray*}
Using the martingale condition, the pricing P.D.E is obtained as
follows:
$$v_t+\frac{1}{2}v_{x_1x_1}B(t,X^0_t,X^1_t)^2+
v_{x_1} A^*(t,X^0_t,X^1_t)+x_1v_{x_0}+x_0v_y+yv_z=rv$$

 For the Asian option based on $Y_t$, where $Y_t=\int_0^t
X_udu$ and $X_t$ is a CARMA($2,1$) process,
%the CARMA($2,1$) processes,
 the following set of S.D.E holds in a similar fashion except that the
 moving average coefficients lead to different representation of
 $Y_t$:

\begin{eqnarray*}
dZ_t&=&Y_tdt\\
 dY_t&=&(b_0X^1_t+b_1X^0_t)dt\\
  dX^0_t&=&X^1_tdt\\
 dX^1_t&=&A^*(t,X^0_t,X^1_t)dt+B(t,X^0_t,X^1_t)dW_t\\
\end{eqnarray*}
This group of differential equations defines a four dimensional
Markovian process. The value of derivative contract will be
$de^{-rt}v(t,x_1,x_0,y,z)$. By Ito'lemma, it is subject to the
following dynamics:
\begin{eqnarray*}
%dv(t,x_1,x_0,y,z)&=&v_t dt+v_{x_1} d{x_1}+\frac{1}{2}v_{x_1x_1}d<x_1>+v_{x_0}dx_0+v_ydy+v_zdz\\
dv(t,x_1,x_0,y,z)&=&v_t dt+v_{x_1} d{x_1}+\frac{1}{2}v_{x_1x_1}d<x_1>+v_{x_0}dx_0+v_ydy+v_zdz\\
dP&=&e^{-rt}(-rv+v_t+\frac{1}{2}v_{x_1x_1}B(t,X^0_t,X^1_t)^2+
v_{x_1}
A^*(t,X^0_t,X^1_t)\\&&+x_1v_{x_0}+b_0x_1v_y+b_1x_0v_y+yv_z)dt+
v_{x_1}
B(t,X^0_t,X^1_t)e^{-rt}dW_t\\
\end{eqnarray*}

Using the martingale condition, the pricing P.D.E is:
$$v_t+\frac{1}{2}v_{x_1x_1}B(t,X^0_t,X^1_t)^2+ v_{x_1}
A^*(t,X^0_t,X^1_t)+x_1v_{x_0}+b_0x_1v_y+b_1x_0v_y+yv_z=rv$$
 The derivative prices for higher order CARMA($p,q$) processes with $p > q$ can be
calculated analytically using this methodology except that
differences in the group of S.D.Es lead to corresponding changes in
the P.D.E terms, which is summarized in the following theorem.
%Finally, for the process where we have formal derivatives ARMA(1,1)
%and ARIMA(1,1,1), Monte Carlo methods can be used within the risk
%neutral framework.

\begin{thm}
If the asset price process is subject to $Y_t$, where $Y_t=\int_0^t
X_udu$ and $X_t$ is a CARMA($p,q$) process
% a
%CARMA($p,1,q$) process $Y_t$
 with $p >
q$, the Asian type call option based on it can be priced via the
following P.D.E:

%$$v_t+A(t,X_t)v+\sum_{i=0}^{q}b_ix_{q-i}v_y+yv_z+\frac{1}{2} B(t,X_t)^2 v_{xx}=rv$$
%$0\leq t<T$,$x_0 \in R$,$\ldots$,$x_{p-1} \in R$,$y \in R$,$z \in R$
%\begin{eqnarray*}
%%v_t+\frac{1}{2}v_{x_{p-1}x_{p-1}}B(t,X^0_t,X^1_t)^2+
%%v_{x_{p-1}} A^*(t,X^0_t,X^1_t)\\
%%+x_{p-1}v_{x_{p-2}}+\cdots+ x_1v_{x_0} +
%%\sum_{i=0}^{q-1}b_ix_{q-i}v_y +yv_z=rv
%v_t+\frac{1}{2}v_{x_{p-1}x_{p-1}}B(t,X^0_t,\cdots,X^{p-1}_t)^2+
%v_{x_{p-1}} A^*(t,X^0_t,\cdots,X^{p-1}_t)
%+\sum_{i=0}^{p-2}x_{i+1}v_{x_{i}}+ \sum_{i=0}^{q}b_ix_{q-i}v_y
%+yv_z=rv
%\end{eqnarray*}

\begin{eqnarray*}
%v_t+\frac{1}{2}v_{x_{p-1}x_{p-1}}B(t,X^0_t,X^1_t)^2+
%v_{x_{p-1}} A^*(t,X^0_t,X^1_t)\\
%+x_{p-1}v_{x_{p-2}}+\cdots+ x_1v_{x_0} +
%\sum_{i=0}^{q-1}b_ix_{q-i}v_y +yv_z=rv
&&v_t+\frac{1}{2}v_{x_{p-1}x_{p-1}}B(t,X^0_t,\cdots,X^{p-1}_t)^2+
v_{x_{p-1}} A^*(t,X^0_t,\cdots,X^{p-1}_t)
\\&&+\sum_{i=0}^{p-2}x_{i+1}v_{x_{i}}+\sum_{i=0}^{q}b_ix_{q-i}v_y
+yv_z=rv
\end{eqnarray*}

where $v=v(t,x_0,\ldots,x_{p-1},y,z)$
 with boundary conditions

%$$lim_{y\rightarrow -\infty}v(t,0,\cdots,0,y,z)=e^{-r(T-t)}(z-K)^+, 0\ leq t < T, z \in R$$
$$lim_{z\rightarrow -\infty} v(t,x_0,\ldots,x_{p-1},y,z)=0, 0 \leq t < T, x_0 \in R,\ldots, x_{p-1}\in R, y \in R$$
$$ v(T,x_0,\ldots,x_{p-1},y,z)=(z-K)^+, x_0 \geq 0, \ldots, x_{p-1}\in R, y \in R, z \in R$$
$$v(t,0,\cdots,0,0,z)=e^{-r(T-t)}(z-K)^+, 0 \leq t < T, z \in R$$
\end{thm}

proof:

Consider the following set of S.D.E
\begin{eqnarray*}
dZ_t&=&Y_tdt\\
 dY_t&=&\sum_{i=0}^{q}b_iX^{q-i}_t dt\\
  dX^0_t&=&X^1_tdt\\
  &&\cdots\\
    dX^{p-2}_t&=&X^{p-1}_tdt\\
 dX^{p-1}_t&=&A^*(t,X^0_t,\cdots,X^{p-1}_t)dt+B(t,X^0_t,\cdots,X^{p-1}_t) dW_t\\
\end{eqnarray*}

This group of differential equation defines a $p+2$ dimensional
Markovian process. The value of derivative contract will be
$de^{-rt}v(t,x_{p-1},\cdots ,x_0,y,z)$. By Ito'lemma, it is subject
to the following dynamics:
\begin{eqnarray*}
%dv(t,x_1,x_0,y,z)&=&v_t dt+v_{x_1} d{x_1}+\frac{1}{2}v_{x_1x_1}d<x_1>+v_{x_0}dx_0+v_ydy+v_zdz\\
dv(t,x_{p-1},\cdots ,x_0,y,z)&=&v_t dt+v_{x_{p-1}} d{x_{p-1}}+\frac{1}{2}v_{x_{p-1}x_{p-1}}d<x_{p-1}>\\
&&+v_{x_{p-2}}dx_{p-2}+\cdots+v_{x_0}dx_0+v_ydy+v_zdz\\
dP&=&e^{-rt}(-rv+v_t+\frac{1}{2}v_{x_{p-1}x_{p-1}}B(t,X^0_t,\cdots,X^{p-1}_t)^2
\\&&+v_{x_{p-1}} A^*(t,X^0_t,\cdots,X^{p-1}_t)
+x_{p-1}v_{x_{p-2}}+\cdots+ x_1v_{x_0}
\\&&+\sum_{i=0}^{q}b_ix_{q-i}v_y
+yv_z)dt+ v_{x_{p-1}}B(t,X^0_t,\cdots,X^{p-1}_t)e^{-rt}dW_t\\
\end{eqnarray*}

Using the martingale condition, the pricing P.D.E is:
\begin{eqnarray*}
%v_t+\frac{1}{2}v_{x_{p-1}x_{p-1}}B(t,X^0_t,X^1_t)^2+
%v_{x_{p-1}} A^*(t,X^0_t,X^1_t)\\
%+x_{p-1}v_{x_{p-2}}+\cdots+ x_1v_{x_0} +
%\sum_{i=0}^{q-1}b_ix_{q-i}v_y +yv_z=rv
&&v_t+\frac{1}{2}v_{x_{p-1}x_{p-1}}B(t,X^0_t,\cdots,X^{p-1}_t)^2+
v_{x_{p-1}} A^*(t,X^0_t,\cdots,X^{p-1}_t)
\\&&+\sum_{i=0}^{p-2}x_{i+1}v_{x_{i}}+\sum_{i=0}^{q}b_ix_{q-i}v_y
+yv_z=rv
\end{eqnarray*}

For the boundary conditions, as $Y_t$ is obtained by removing trends
and seasonal adjustments from the original travel time series, it
can be negative. All the state variables corresponding to the
CARMA($p,q$) process can be in $R$.

If $Y(t)$ approaches $-\infty$, then the probability that the call
expires in the money approaches zero and the option price approaches
zero. This leads to the first boundary condition. The second
boundary condition is just the payoff for the call option at time
$T$. The third boundary condition follows the discounted payoff of
Asian call option from $T$ to $t$.

 Q.E.D

In order to numerically solve the equation in the theorems above, it
would normally be necessary to also specify the behavior of $v$ as
all variables approaches $+\infty$ or $-\infty$, which can be
different to each case. Moreover, the prices of put options and
other derivatives based on the travel time process can be computed
in a similar fashion.

To connect this risk neutral representation with typical hedging
strategies, the derivation in the previous section is applied to
$Z_t$: Suppose the two derivatives which are both derived on $Z_t$
but have two different payoff functions: $F$ and $G$. A portfolio is
defined $P=\alpha F +\beta G$ of $F$ and $G$ with $\alpha+\beta=1$.
If the portfolio is risk neutral, then its value should increase at
the same rate as a risk-free rate. The following P.D.E can be then
derived, defining $\theta_t$ as the market value of risk.
\begin{eqnarray*}
&&v_t+\frac{1}{2}v_{x_{p-1}x_{p-1}}B(t,X^0_t,\cdots,X^{p-1}_t)^2+
v_{x_{p-1}} (A(t,X^0_t,\cdots,X^{p-1}_t)-\theta_t
B(t,X^0_t,\cdots,X^{p-1}_t))
\\&&+\sum_{i=0}^{p-2}x_{i+1}v_{x_{i}}+\sum_{i=0}^{q}b_ix_{q-i}v_y
+yv_z=rv
\end{eqnarray*}

This P.D.E incorporates more explicitly the market value of risk and
corresponding hedging strategy, while it maintains similar
theoretical properties as the general P.D.E in the theorem above. As
discussed in the previous section, different hedging strategies may
introduce different risk neutral measures in an incomplete market.

 In summary, this paper introduces travel time derivatives as
an innovative value pricing scheme, an effective hedging tool
against risk due to bad quality of traffic service and a new
financial instrument to diversify portfolio risk. The market
participants are mainly travelers and business whose businesses may
be influenced by the traffic system and typical financial
derivatives such as futures and options can be derived based on
travel time. Ornstein - Uhlenbeck process and more generally, the
continuous time auto regression moving average (CARMA) models are
used to model travel time while risk neutral pricing principle under
incomplete market conditions is used to price such products; both
explicit P.D.E solutions and Monte Carlo methods are used to obtain
the numerical asset prices. The analysis of financial derivative
based on travel time extends the literature of derivative pricing
based on non-tradable assets to new disciplines and leads to
enormous research opportunities.

%\bibliographystyle{agsm}
%\bibliography{1,2,3}

\section{Acknowledgement}

The authors would like to thank Rachel Blum for her help on grammar and phrasing.

%\section{Reference}
%Bibliography
\bibliographystyle{unsrt}  
\bibliography{references}

\end{document}